\documentclass[letter,nocrop]{bioinfo_arxiv} 
\usepackage{dsfont}
\usepackage{physics}
\usepackage{algorithm}
\usepackage{algpseudocode}
\usepackage{hyperref}

\copyrightyear{2021} \pubyear{2021}

\access{Advance Access Publication Date: Day Month Year}

\appnotes{Research article}

\begin{document}

\firstpage{1}

\subtitle{Subject Section}

\title[Topological Approximate Bayesian Computation]{Topological Approximate Bayesian Computation for Parameter Inference of an Angiogenesis Model}

\author[Thorne \textit{et~al}.]{Thomas Thorne\,$^{\text{\sfb 1,}*}$, Paul D W Kirk\,$^{\text{\sfb 2,\sfb 3,\sfb 4}}$ and Heather A Harrington\,$^{\text{\sfb 5,6,}*}$}
\address{$^{\text{\sf 1}}$Department of Computer Science, University of Surrey, Guildford, GU2 7XH, United Kingdom\\
$^{\text{\sf 2}}$MRC Biostatistics Unit, University of Cambridge, Cambridge, CB2 0SR, United Kingdom.
\\
$^{\text{\sf 3}}$Cambridge Institute of Therapeutic Immunology \& Infectious Disease (CITIID), University of Cambridge, Cambridge, CB2 0AW, United Kingdom.\\
$^{\text{\sf 4}}$Cancer Research UK Cambridge Centre, Ovarian Cancer Programme, Cambridge, CB2 0RE, United Kingdom.\\
$^{\text{\sf 5}}$Mathematical Institute, University of Oxford, Oxford, OX2 6GG, United Kingdom.\\
$^{\text{\sf 6}}$Wellcome Centre for Human Genetics, University of Oxford, Oxford, OX3 7BN, United Kingdom.
}

\corresp{$^\ast$To whom correspondence should be addressed.}

\history{Received on XXXXX; revised on XXXXX; accepted on XXXXX}

\editor{Associate Editor: XXXXXXX}

\abstract{\textbf{Motivation} Inferring the parameters of models describing biological systems is an important problem in the reverse engineering of the mechanisms underlying these systems. Much work has focused on parameter inference of stochastic and ordinary differential equation models using Approximate Bayesian Computation (ABC). While there is some recent work on inference in spatial models, this remains an open problem. Simultaneously, advances in topological data analysis (TDA), a field of computational mathematics, have enabled spatial patterns in data to be characterised. \textbf{Results} Here we focus on recent work using topological data analysis to study different regimes of parameter space for a well-studied model of angiogenesis. We propose a method for combining TDA with ABC to infer parameters in the Anderson-Chaplain model of angiogenesis. We demonstrate that this topological approach outperforms ABC approaches that use simpler statistics based on spatial features of the data. This is a first step towards a general framework of spatial parameter inference for biological systems, for which there may be a variety of filtrations, vectorisations, and summary statistics to be considered. \textbf{Availability and Implementation} All code used to produce our results is available as a Snakemake workflow from \url{github.com/tt104/tabc_angio}.}

\maketitle


\section{Introduction}
  

When analysing mathematical models of biological systems, we often aim to reverse engineer the parameters of the model
by fitting to observed data. The Bayesian formalism provides a principled way to perform parameter inference that quantifies our uncertainty in the model parameters \citep[see, for example,][]{kirk2015systems}, but traditionally requires us to be able to write down an analytical function (the likelihood function) that returns the likelihood of a parameter vector given the observed data.

However, for many models of interest, there is no straightforward way to write down the likelihood function associated with the model. This is often due to the intractability of deriving a closed form expression for the model likelihood. In such situations, it may nevertheless be possible to apply a simulation-based inference approach termed {\em Approximate Bayesian Computation}  \citep[ABC; see, for example,][]{sissonHandbookApproximateBayesian2018}, that substitutes a kernel on some statistics of the data for the model likelihood, and evaluates the fit of the model at a given set of parameter values through simulations. For given parameter realisations, the model is simulated, and the statistics of the simulated data compared with the same statistics of the observed data. Informally, regions of parameter space that correspond to simulated data sets whose statistics are ``more similar" to those of the observed data will be associated with higher posterior probability than regions corresponding to simulated data sets with statistics that are ``less similar" (where ``similarity" is quantified using a pre-specified distance function). 

Applying ABC, we can derive an approximate posterior distribution over the model parameters using standard sampling techniques such as rejection sampling. This approximate posterior distribution expresses our uncertainty in the model parameters, given the model and the observed data set. Recently, ABC parameter inference and model selection has been successfully developed for reaction-diffusion models \citep{warne2019using}. However, performing parameter inference for more general spatial models has been largely unexplored.


Topological data analysis (TDA) is a relatively new area of computational mathematics that quantifies the shape of data by computing topological properties of the data. There are various approaches of topological inference, e.g., level sets or mode clusters \citep{wassermanTopologicalDataAnalysis2018}. The most prominent algorithm in TDA is persistent homology \citep[PH;][]{carlssonTopologyData2009a,edelsbrunnerComputationalTopologyIntroduction2010}. PH takes in data and a metric, and outputs topological features (e.g., connected components and loops) and their persistence across different scales of the data. The computation  crucially depends on the choice of filtration, which is a nested sequence of spaces built on the data, that is indexed by a scale parameter \citep{edelsbrunnerComputationalTopologyIntroduction2010,ghristHomologicalAlgebraData2018a}. 
There are many software implementations for persistent homology \citep{Otter2017}; however, the software used is often selected based on the types of filtrations available within it. 
The choice of filtration for applications is an active area of research, and there is no one-size-fits-all filtration for biological applications \citep{stolz-pretzer_global_2019}. The persistence of the topological features as well as where topological features appear and die in the filtration may provide insight into biological processes and models.


In previous work with spatial models of biological processes \citep{murrayMathematicalBiologyII2003}, TDA has been applied to test for spatial randomness \citep{robinsPrincipalComponentAnalysis2016}, automatically detect zebra-fish patterns  \citep{mcguirl2020topological}, characterise immune cell infiltration by changes in a chemotaxis parameter \citep{Vipond2021}, and  cluster parameter regimes for angiogenesis \citep{nardiniTopologicalDataAnalysis2021}.  Now we wish to address the inverse problem of recovering model parameters given some observed data, in the Bayesian formalism. ABC enables us to perform parameter inference in a statistical model on the basis of data summaries, even when there is no clear way to define a likelihood function for the model. One key challenge in ABC is the choice of summary statistic, as the statistic must capture the relevant information about the model parameters in the data to allow the parameters to be learnt. Here we show that TDA provides informative data summaries that enable parameter inference to be performed successfully in a spatial model.  In particular, we consider as a case study the Anderson-Chaplain model of angiogenesis \citep{andersonContinuousDiscreteMathematical1998}.


In previous work in the literature \citet{maroulasBayesianFrameworkPersistent2020} model persistence diagrams as Poisson point processes and use this to allow a posterior to be inferred on a persistence diagram given some observed data and a suitable prior. This allows a posterior on topological features to be defined, and a scheme for performing Bayesian classification is developed, but it does not consider the case of performing inference on a parametric model, given an observed set of topological features.

In \citet{sgouralisBayesianTopologicalFramework2017}, Bayesian inference is applied in the processing of the data, but not in a topological context or for parameter inference in the model of interest. Instead various performance measures are evaluated for a small set of selected parameter combinations, not considering a distribution over parameters or a Bayesian posterior.

In this paper we first describe the model and data generation process applied, before describing TDA and ABC in general terms, and their specific application to the Anderson-Chaplain model. We demonstrate our suggested approach for parameter inference on simulated data from the Anderson-Chaplain model and compare the outputs to the results produced by other non-topological statistics.

\section{Model Data}

The Anderson-Chaplain model \citep{andersonContinuousDiscreteMathematical1998} is a well-studied spatio-temporal model of angiogenesis. Angiogenesis is the growth of new blood vessels from pre-existing vasculature. The model combines a system of partial differential reaction equations with discrete dynamics. The model considers production and consumption of fibronectin, the secretion of tumour angiogenic factors (TAF) from a tumour, and new vasculature forms from endothelial tip cells in response to gradients of fibronectin and TAF.

The Anderson-Chaplain model of angiogenesis has two key parameters, $\rho$ and $\chi$, coefficients for haptotaxis and chemotaxis respectively. These determine the relative contribution of fibronectin driven haptotaxis and TAF driven chemotaxis to the movement of tip cells in the model. Other parameters determine the dynamics of the distribution of fibronectin and TAF, and we keep these fixed as in \citet{nardiniTopologicalDataAnalysis2021}.

Data were generated by simulating the Anderson-Chaplain model on a two-dimensional square lattice of resolution $201$ by $201$ \citep[as in][]{andersonContinuousDiscreteMathematical1998} using the implementation provided in \citet{nardiniTopologicalDataAnalysis2021}, with a linear chemoattractant distribution that increases with the coordinate along the $x$ axis. This produces sets of binary images (see figure \ref{fig:posterior}) which are then further processed using the methods described below.

\section{Methods}

\subsection{Topological Data Analysis}

To characterise the $k$ dimensional features of a topological space $X$ we can consider the homology group in dimension $k$, $H_k(X)$, composed of elements that intuitively correspond to equivalence classes of cycles that can be continuously deformed into one another on $X$. In dimension one, the generators of the homology group correspond to one dimensional holes in $X$, or loops, while in dimension zero the generators of the homology group correspond to the connected components of $X$.

The topological spaces we are interested in can be represented using finite sets of simplices 
known as simplicial complexes $K$ that are constructed by joining together individual simplices, potentially of different dimension, and are closed under the operation of taking faces. A zero dimensional simplex corresponds to a single vertex, a one dimensional simplex an edge, and a two dimensional simplex a triangle. Given a real valued function on $K$ we can define a filtration as a sequence of homology groups in a given dimension $k$, with homomorphisms induced by inclusion

\begin{equation}
    0 = H_k(K_{a_0}) \rightarrow H_k(K_{a_1}) \rightarrow \ldots \rightarrow H_k(K_{a_n}) = H_k(K)
\end{equation}
where $K_a=f^{-1}(-\infty,a]$ and $a_0<a_1<\ldots<a_n$, and $K_{a_i}\subseteq K_{a_j}$ for $i<j$. Persistent homology then tracks the birth and death of elements of the homology groups as $a$ varies. 
By choosing an appropriate definition of the simplicial complex and filtration built from the data, persistent homology 
can provide information about the topological features in data.

We build the simplicial complex and filtration from the final timepoint of model simulation data following \citet{nardiniTopologicalDataAnalysis2021}. All cells in the two-dimensional square lattice that have vasculature present are assigned a value of one, and zero elsewhere. The centroid of each non-zero cell is  a $0$-simplex. The simplicial complex is built on these $0$-simplices based on so-called Moore neighborhoods: if any of the eight cells surrounding a vertex are also nonzero, then we connect them via $1$-simplices (edges) for two points pairwise connected, or $2$-simplices for three points pairwise connected by an edge. The union of these simplices form a \emph{simplicial complex}.  There are different ways to study vascular data at multiple scales using filtrations ~\citep{bendich2016persistent,stolz2020multiscale}.  Here, we construct sequences of filtered simplicial complexes using a sweeping plane filtration \cite{bendich2016persistent,nardiniTopologicalDataAnalysis2021}. In the sweeping plane filtration, we move a vertical line from left to right across the 2D lattice domain and include simplices in the filtration only to the left of this line. This filtration can be considered a sublevel set filtration corresponding to a height function $h: X \rightarrow \mathbb{R}$ on this simplicial complex.

\subsection{Approximate Bayesian Computation}

In Bayesian inference we aim to derive the posterior distribution of the parameters of a model given some observed data. To do so we first define a prior distribution on the model parameters, treating them as random variables. This describes our belief in the distribution of the parameters before having observed any data. We then perform a so called \textit{Bayesian update} of the model having observed some data. This is done using the likelihood of the observed data given the model and parameters. From this we arrive at a posterior distribution that describes the conditional distribution of the parameters given the observed data. If we denote the model parameters by $\theta$, and the data by $x$, we can first write the prior as $p(\theta)$, and the likelihood of the data as $p(x|\theta)$. In the Bayesian framework we apply Bayes rule to update the prior distribution having observed the data, giving us the posterior distribution as

\begin{equation}
    p(\theta|x) = \frac{p(x|\theta)p(\theta)}{p(x)},
\end{equation}

where $p(x)$ is known as the evidence or marginal likelihood, and plays a key role in Bayesian model selection. Evaluation of the marginal likelihood is often computationally expensive or intractable.  However, in many settings (e.g. when sampling from the posterior using Markov chain Monte Carlo techniques), it is sufficient to be able to write down the posterior up to proportionality
\begin{equation}
    p(\theta|x) \propto p(x|\theta)p(\theta).
\end{equation}

This approach relies on the ability to calculate both the prior of the parameters $p(\theta)$, which is generally tractable, and the likelihood $p(x|\theta)$. However in many models of interest it is not tractable or not possible to directly evaluate $p(x|\theta)$, for example in population genetics \citep{beaumontApproximateBayesianComputation2002}, random graph models \citep{thorneGraphSpectralAnalysis2012a} and some models of dynamical systems \citep{toniApproximateBayesianComputation2009a,liepe2014framework}. To allow us to perform Bayesian inference in these situations, an approach named Approximate Bayesian Computation (ABC) was developed, based on initial work in \citet{fuEstimatingAgeCommon1997} and \citet{tavareInferringCoalescenceTimes1997}, developed further in \citet{beaumontApproximateBayesianComputation2002} and \citet{marjoramMarkovChainMonte2003a}, and expanded in many works, see e.g. \citet{sissonSequentialMonteCarlo2007a,toniApproximateBayesianComputation2009a,beaumontAdaptiveApproximateBayesian2009,delmoralAdaptiveSequentialMonte2012,prangleRareEventApproach2018}.

In an ABC framework, we rely on the observation that given the ability to sample realisations $y$ from $p(x|\theta)$, we can rewrite the posterior as

\begin{equation}
    p(\theta|x) = \int p(\theta,y|x)\dd{y} ,
\end{equation}

where

\begin{equation}
    p(\theta,y|x) = \frac{\mathds{1}(x=y)p(y|\theta)p(\theta)}{p(x)},
\end{equation}

and by relaxing this to

\begin{equation}
    p(\theta,y|x) \approx \frac{\mathds{1}(D(x,y)<\epsilon)p(y|\theta)p(\theta)}{p(x)},
\end{equation}

we can generate samples from an approximate posterior (which we shall refer to as the {\em ABC posterior}) by using a suitably small $\epsilon$ in Algorithm~\ref{a:abc}. Often when applying the rejection algorithm we fix the number of samples $S$ and select $\epsilon$ such that the set of samples $\hat{\theta_s}$ with $d_s<\epsilon$ is some fraction $\alpha S$.

\begin{algorithm}[h]
\caption{ABC rejection sampler algorithm}\label{a:abc}
\begin{algorithmic}[1]
\For{$s \in 1,\ldots,S$}
\State{Sample $\hat{\theta_s}\sim p(\theta)$}
\State{Simulate $y\sim p(y|\hat{\theta_s})$}
\State{Calculate $d_s\leftarrow D(g(y),g(x))$}
\EndFor
\State{Return samples $\hat{\theta_s}$ where $d_s<\epsilon$}
\end{algorithmic}
\end{algorithm}

The ABC rejection sampler algorithm requires us to define a distance on the data, $D(x,y)$, and in some cases this may itself be intractable. It is then possible to substitute a summary statistic of the data, $g(x)$ in place of the data itself, leading to a distance on these summary statistics $D(g(x),g(y))$ being considered. In the case where $g$ is a \textit{sufficient statistic} for the model, as $\epsilon \rightarrow 0$ this will be equivalent to applying a distance on the $x$ and $y$ themselves. Often this is not the case, and this is another avenue through which ABC produces an approximation to the posterior rather than a true evaluation of the posterior itself.

\subsection{Topological statistics for Approximate Bayesian Computation}

In previous work \citet{nardiniTopologicalDataAnalysis2021} applied topological statistics of simulated data (2-D binary images) to quantify different regimes in the parameter space of the Anderson-Chaplain model of angiogenesis. By constructing simplicial complexes from the output data of a spatial model, and using the same filtration as \citet{nardiniTopologicalDataAnalysis2021}, PH can be applied to describe the presence of topological features in the simulated data.

In some cases when calculating the persistence of the topological features of a filtration, it is possible for some features to persist indefinitely, so that their death in the filtration is represented as $+\infty$. In our application, this causes information about certain topological features to be lost, for example loops and some connected components, as although we know when they are born in the filtration, we have no measure of their extent. 
 For this reason, \citet{nardiniTopologicalDataAnalysis2021} computed persistence of a left to right sweeping plane filtration and right to left sweeping plane filtration of the simplicial complex built from the simulated model data (see \citet{nardiniTopologicalDataAnalysis2021} for details).
By viewing the left to right filtration as a sublevel set filtration and the right to left filtration as a superlevel set filtration, more information (e.g., only finite bars that capture the extent of topological features) can be extracted as a consequence of duality and symmetry theorems \citep{cohen-steinerExtendingPersistenceUsing2009}. 

\subsection{Extended persistence}

Here we propose a more elegant solution that applies the extended persistence of \citet{cohen-steinerExtendingPersistenceUsing2009}, which forces all topological features to be of finite length.
Extended persistence was developed to study cavities and protrusions in protein docking \citep{agarwal2006extreme,cohen-steinerExtendingPersistenceUsing2009}. Since then, \citet{yim2021optimization} optimised spectral wavelets for graph classification using extended persistence, and extended a differentiability result for ordinary persistence to extended persistence. 

In standard persistence, the sublevel sets $X_a=f^{-1}(-\infty,a]$ of the manifold $X$ are nested and PH is defined through the corresponding linear sequence of homology groups. 
In extended persistence, we compute the homology of the sublevel sets, as well as the relative homology with respect to the superlevel sets $X^a=f^{-1}[a,\infty)$. 
This is motivated by the fact that the relative homology group $H_k(X,X^a)$ of dimension $k$ is isomorphic to the cohomology group of $X_a$ of dimension $dim-k$, denoted $H^{dim-k}(X_a)$, where $dim$ is the dimension of the manifold $X_a$ \citep{cohen-steinerExtendingPersistenceUsing2009}.

For a set of values $a_0,\ldots,a_n$ that bound and fit between the critical points of $f$, the extended persistence in dimension $k$ is defined as the persistence of the homology groups and relative homology groups as
\begin{align}
    0 &= H_k(X_{a_0}) \rightarrow H_k(X_{a_1}) \rightarrow \ldots \rightarrow H_k(X_{a_n}) = H_k(X)\nonumber\\ 
    &H_k(X) = H_k(X,X^{a_n}) \rightarrow \ldots \rightarrow H_k(X,X^{a_0}) = 0   
\end{align}
where $H_k(X,X^a)$ denotes the relative homology group of $X$ and $X^a$ in dimension $k$ \citep{edelsbrunnerComputationalTopologyIntroduction2010}.

This extended persistence can be broken down into multiple components \citep{cohen-steinerExtendingPersistenceUsing2009}, the ordinary part, formed of topological features that are both born and die within the homology groups of the sublevel sets of $X$, the relative part of features that are born and die in the relative homology groups, and the extended part of features that are born in the ordinary homology groups and die in the relative homology groups in the filtration. The birth time $b$ of a feature may be larger than its death time $d$ due to the possibility that the feature dies in the relative homology group $H(X,X^d)$ with $d<b$. The extended part can be further divided into topological features that have $b<d$, termed extended$+$, and those with $d<b$, termed extended$-$.



\subsection{Persistence images}

The output of applying PH to a data set is often represented as a persistence diagram, that for a given dimension $k$ consists of a plot of points $(b,d)$, where $b$ is the time of birth and $d$ is the time of death $d$ of each dimension $k$ topological feature in the filtration. To allow for the straightforward application of methods from machine learning to these diagrams, \citet{JMLR:v18:16-337} developed the concept of a persistence image. This allows a persistence diagram to be represented as a vector in $\mathds{R}^n$, so that for example it can be used in methods such as K-means clustering, as in \citet{nardiniTopologicalDataAnalysis2021}.

To generate the persistence image corresponding to a persistence diagram represented as a multiset of points $(b,d)$, the points are first transformed to give a multiset $B$ of birth and persistence coordinates $(b,d-b)$ (for extended persistence, we require a slightly different formulation -- see below). We note that the persistent image formulation of \citet{JMLR:v18:16-337} ignores all infinite persistent features. A persistence surface in $\mathds{R}^2\rightarrow \mathds{R}$ is then defined as the weighted sum of kernels applied to each birth/persistence coordinate

\begin{equation}
    f(x,y) = \sum_{(b,p)\in B} g(b,p)h(x,y;b,p).
    \label{e:persist}
\end{equation}

From the persistence surface defined in eqn. \ref{e:persist}, an $m\cross m$ array of values is created by discretizing $f(x,y)$ into an $m$ by $m$ grid in a suitable range. This array can then by flattened to give a vector in $\mathds{R}^{m^2}$. As in \citet{JMLR:v18:16-337} we apply a Gaussian kernel for $h$ with mean $\mu=(b,p)$ and fixed standard deviation $\sigma$.

We remark that extended persistence only has finite persistence; therefore no information (i.e., the infinite bars in ordinary persistence) is lost in the persistence images for extended persistent homology. 

\subsection{TABC}

We use a set of topological statistics derived from the extended persistence of a filtration over the simplicial complex representing the data as the summary statistics in an ABC framework, in a method we title TABC, to perform topological posterior inference on the Anderson-Chaplain model of angiogenesis. In the TABC methodology, the summary statistics used in ABC are the persistence images in each dimension produced by the by the four components of the extended persistence of a filtration. To allow persistence images to be generated for the extended persistence, in components of the extended persistence with points in the persistence diagram $(b,d)$ with $d<b$, we flip the coordinates to consider instead $(d,b)$, which when transformed into a birth/persistence coordinate then represents the duration of persistence of the feature in the relative part, or the gap between birth in the ordinary homology and death in the relative homology of the feature in the extended$-$ part.
We generate persistence images of dimension $50$ by $50$  with a constant weight function for the persistence surface and the kernel of the persistence images set as a Gaussian distribution with standard deviation $\sigma=1$, as we found this to work well. As the distance metric in the ABC algorithm we applied the Euclidean distance between the statistics. 
In our implementation we use the GUDHI library (\url{http://gudhi.gforge.inria.fr/}) to construct simplicial complexes, generate extended persistence diagrams and produce persistence images (with standard weighting $g=1$).

\subsection{Image-based statistics}
\label{s:im}

For comparison we also consider four statistics based on the binary image data produced by the simulations, that were chosen with the aim of differentiating the different classes of behaviours observed in \citet{nardiniTopologicalDataAnalysis2021}, without overlapping with features that could be considered as topological descriptors (for example numbers of connected components). These statistics are:

\begin{itemize}
\item \textbf{Mean X coordinate} The mean X value of occupied pixels.
\item \textbf{Mean Y coordinate} The mean Y value of occupied pixels.
\item \textbf{Maximum X coordinate} The maximum X value of an occupied pixel.
\item \textbf{Mass} The fraction of occupied pixels.
\end{itemize}

As with the topological statistics, we applied the Euclidean distance between vectors of statistics as the distance in the ABC rejection algorithm.

\section{Results}

\begin{figure*}[!htp]
\centering
\begin{tabular}{ccccc}
Observed&Posterior density&\multicolumn{3}{c}{ABC posterior predictive}\\
\includegraphics[width=0.12\textwidth]{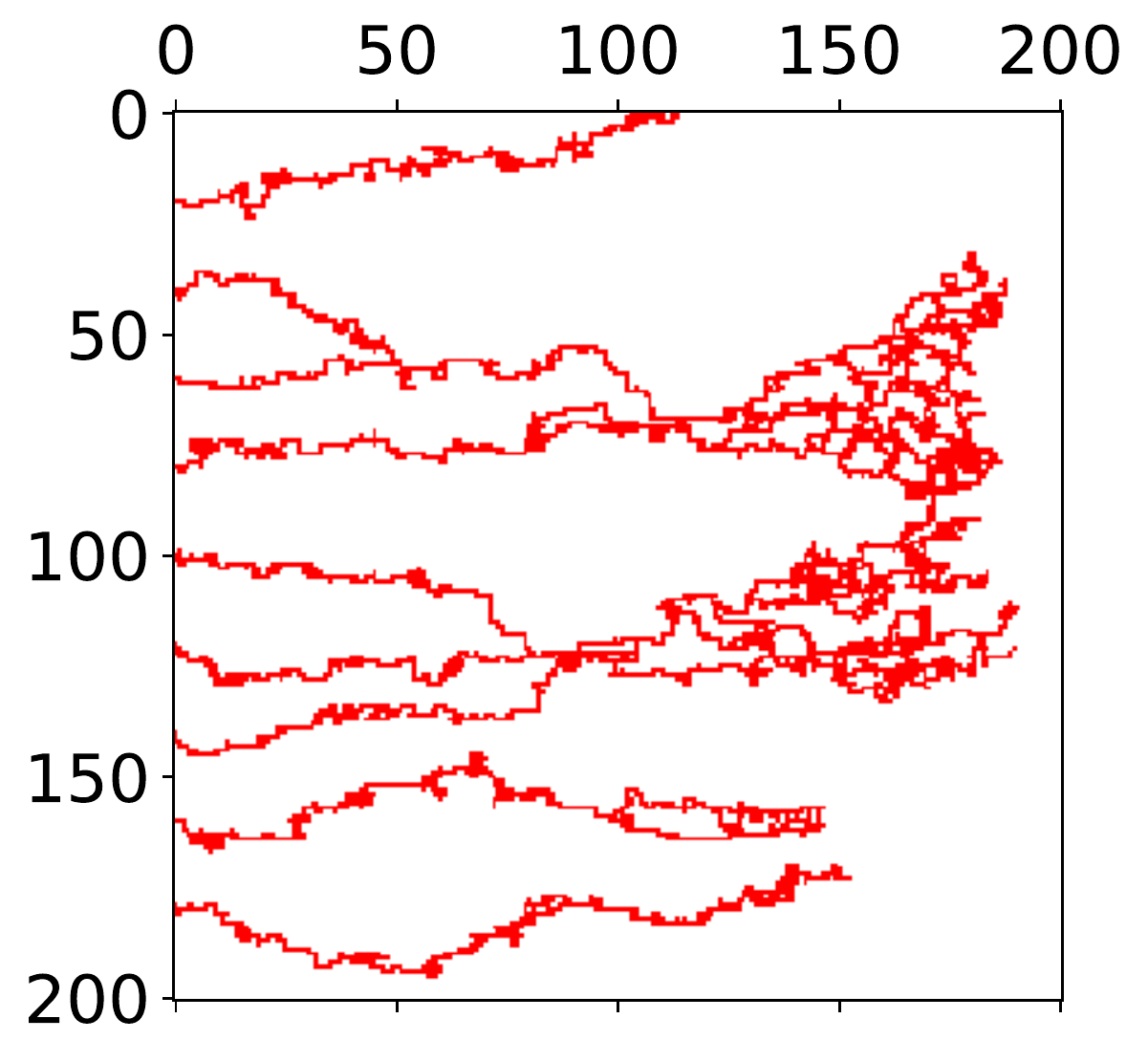}&
\includegraphics[width=0.11\textwidth]{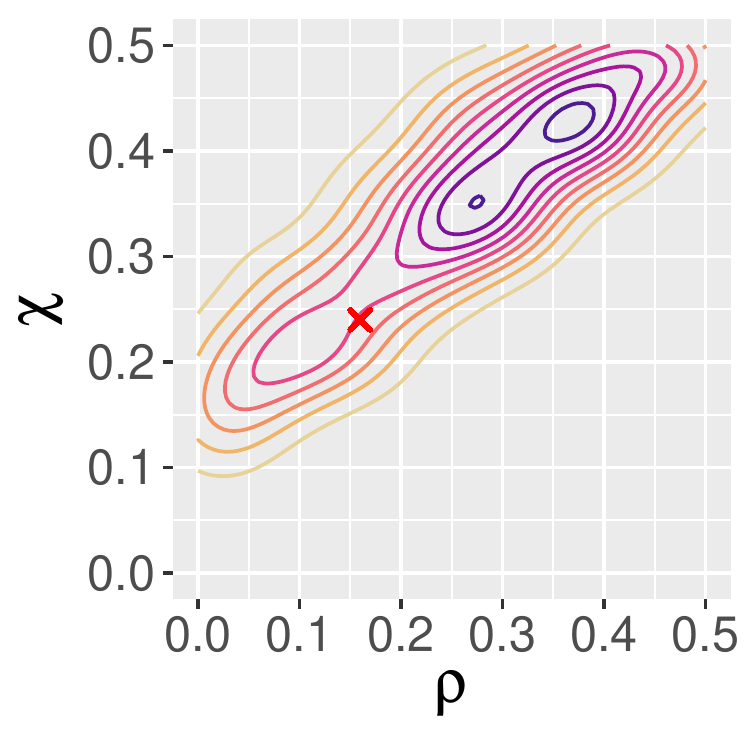}&
\includegraphics[width=0.12\textwidth]{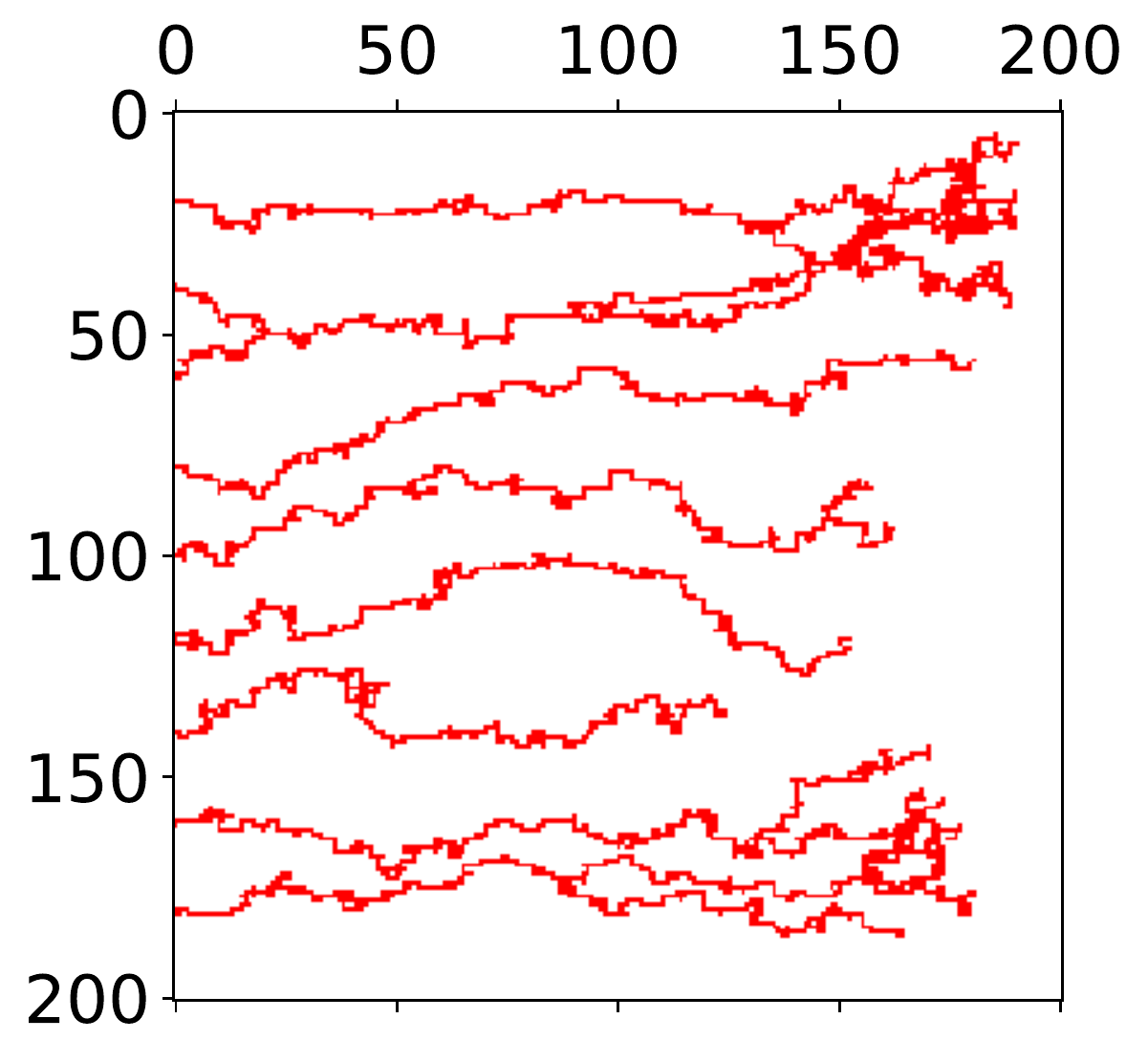}&
\includegraphics[width=0.12\textwidth]{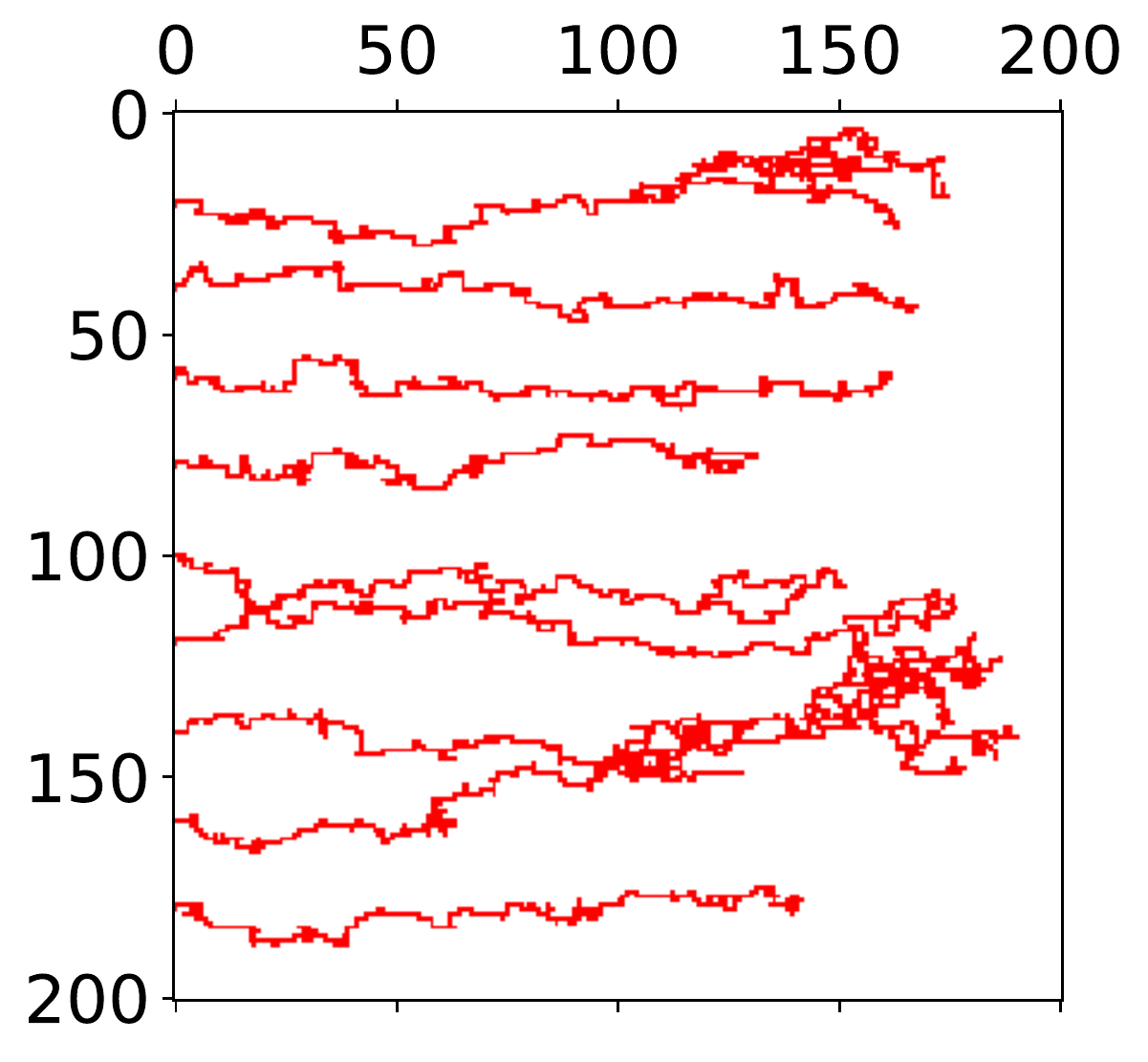}&
\includegraphics[width=0.12\textwidth]{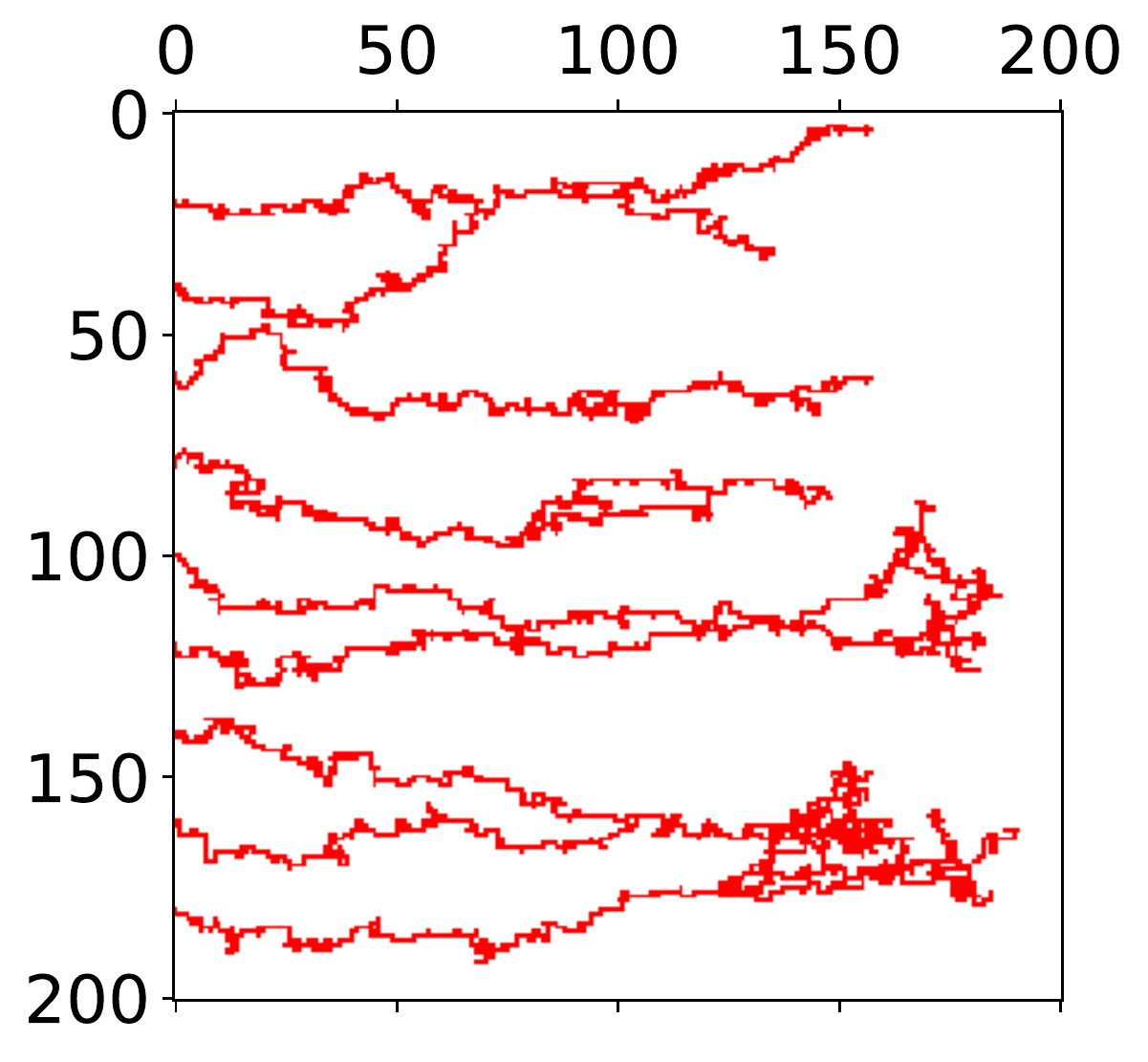}\\

\includegraphics[width=0.12\textwidth]{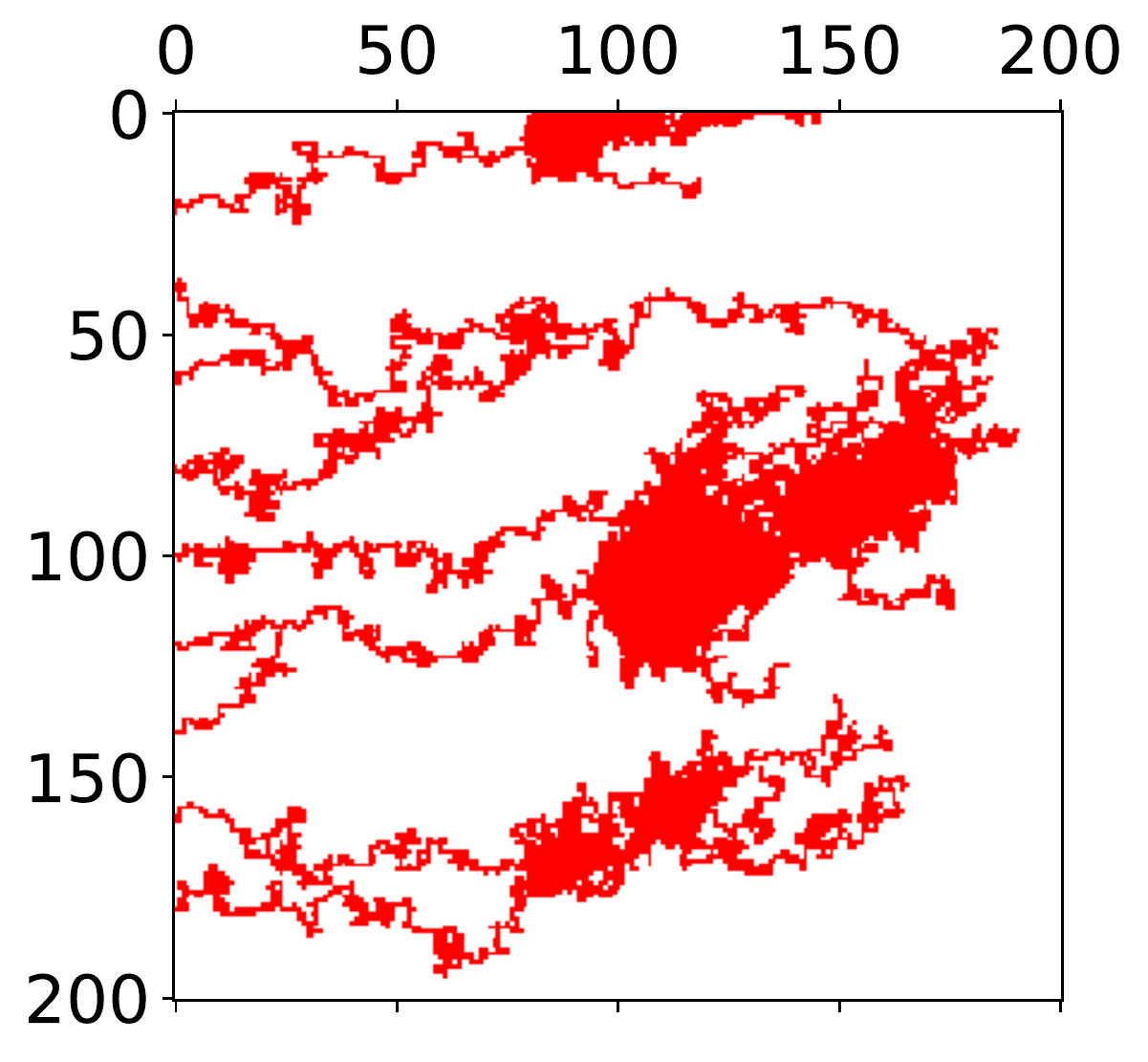}&
\includegraphics[width=0.11\textwidth]{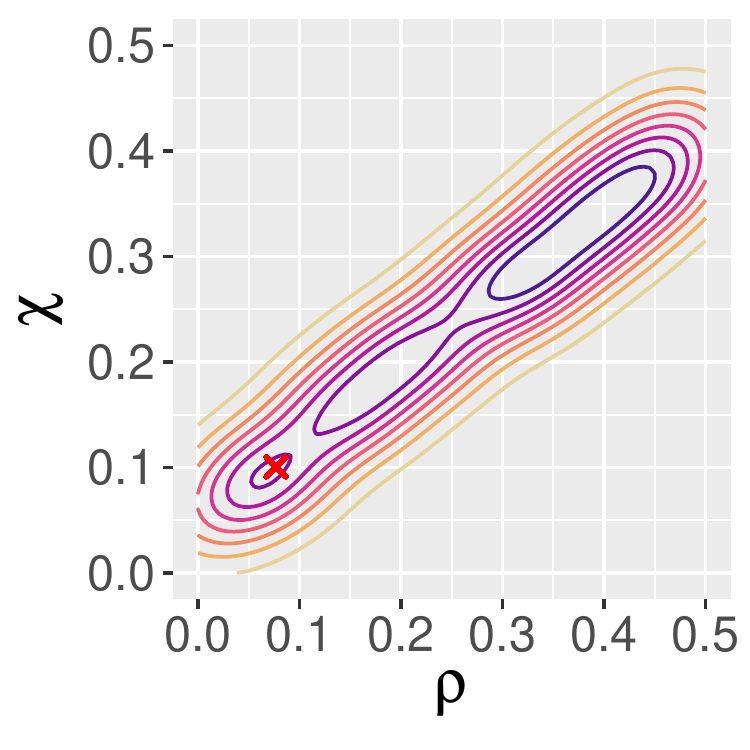}&
\includegraphics[width=0.12\textwidth]{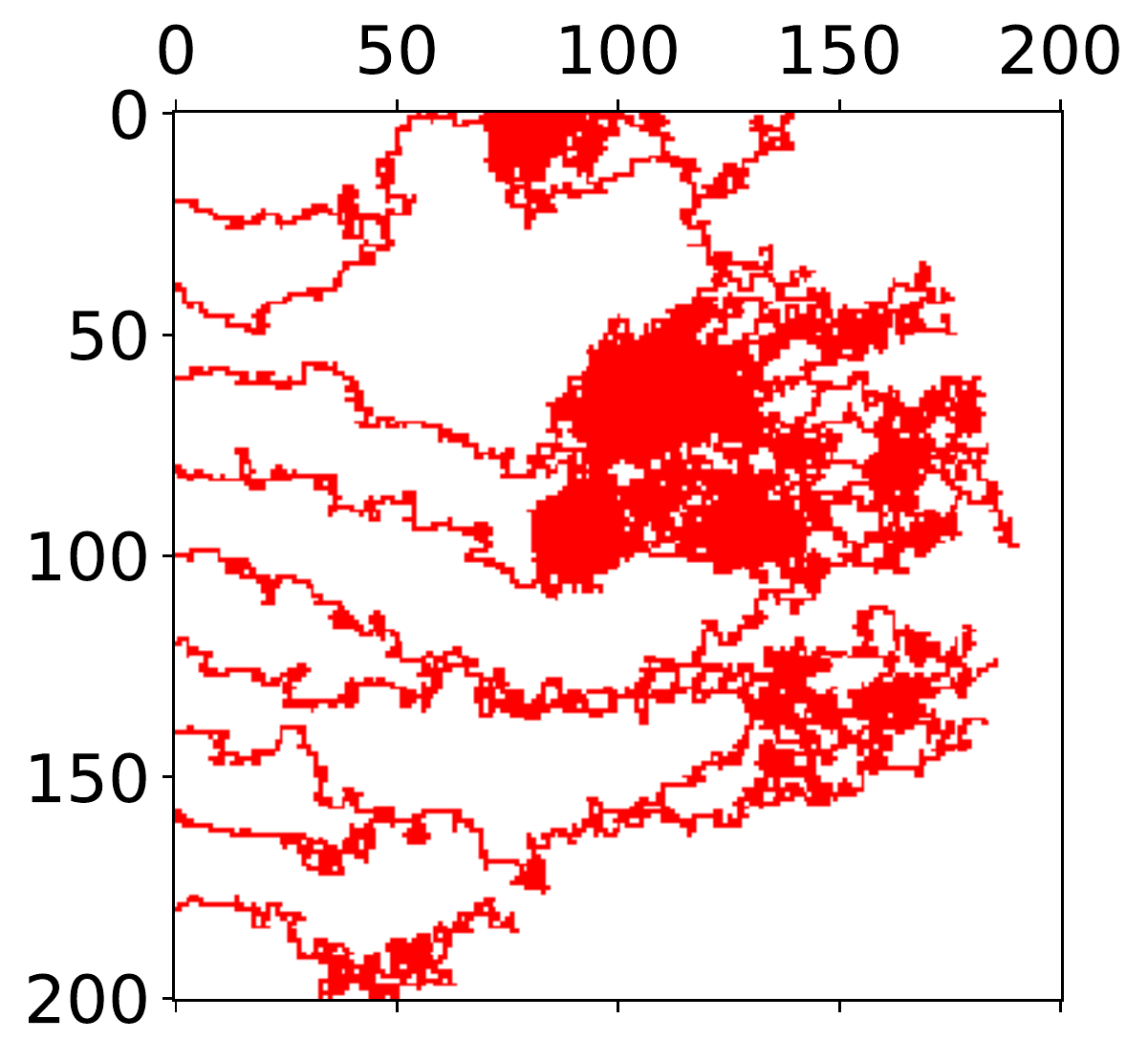}&
\includegraphics[width=0.12\textwidth]{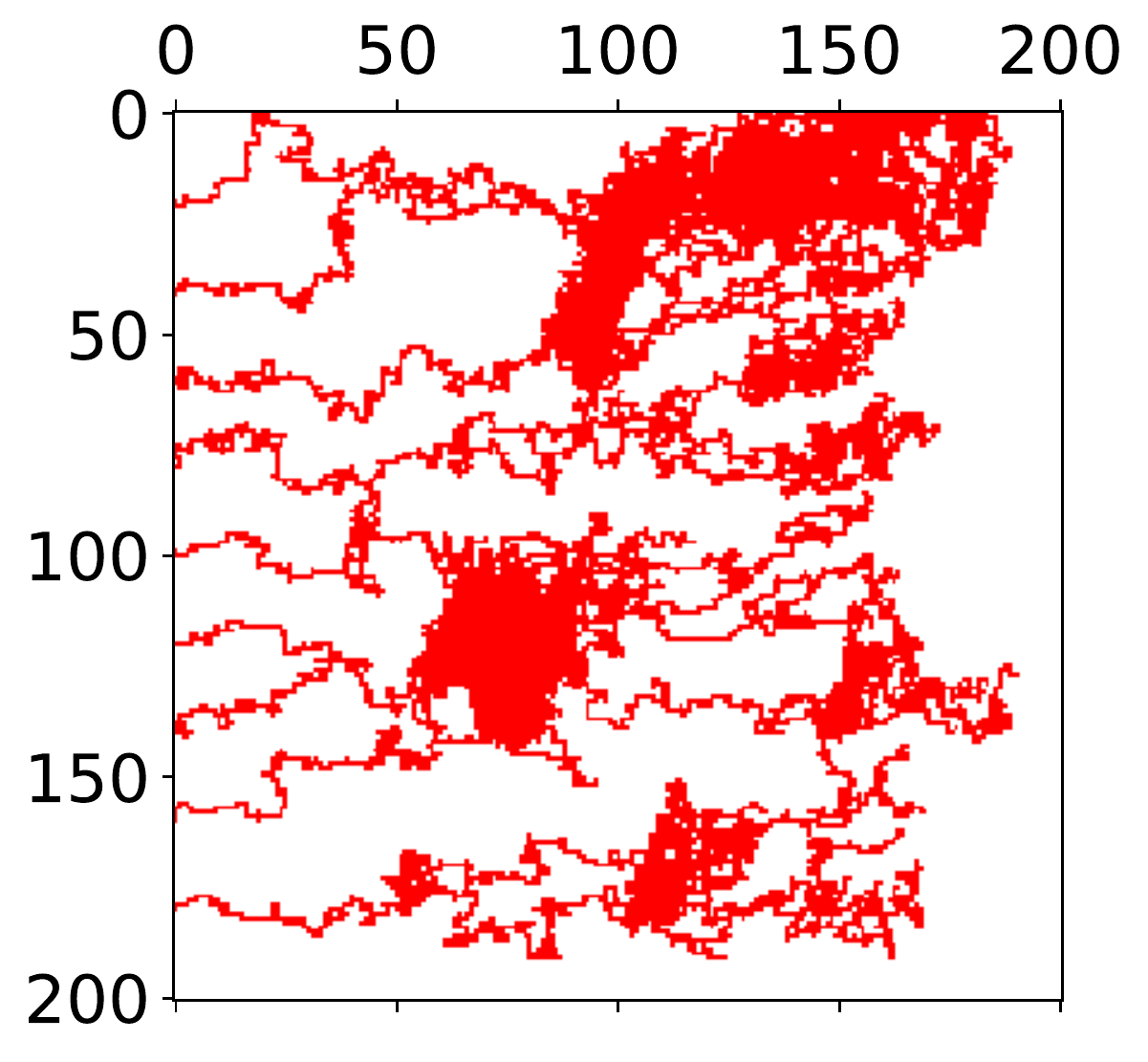}&
\includegraphics[width=0.12\textwidth]{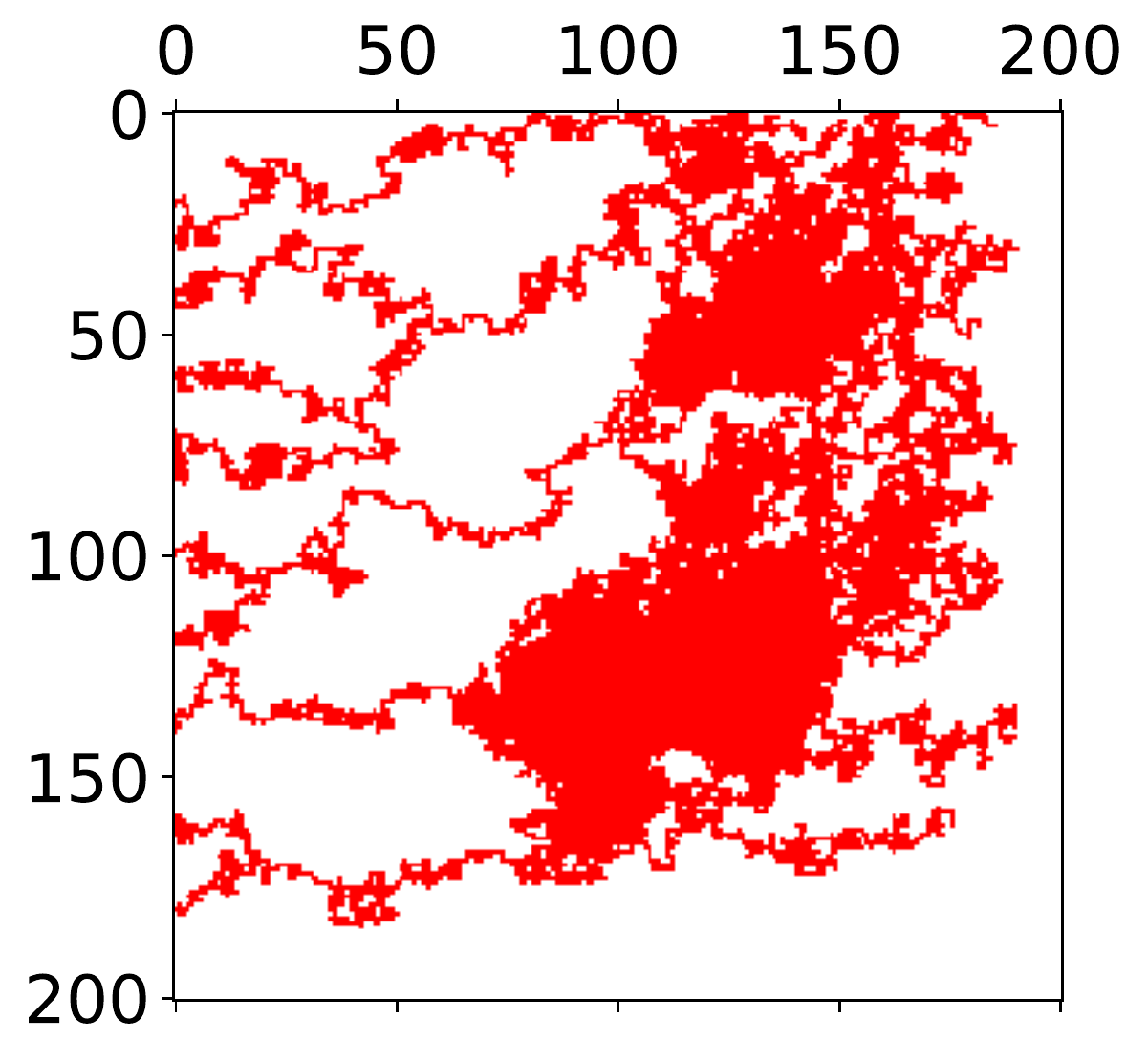}\\

\includegraphics[width=0.12\textwidth]{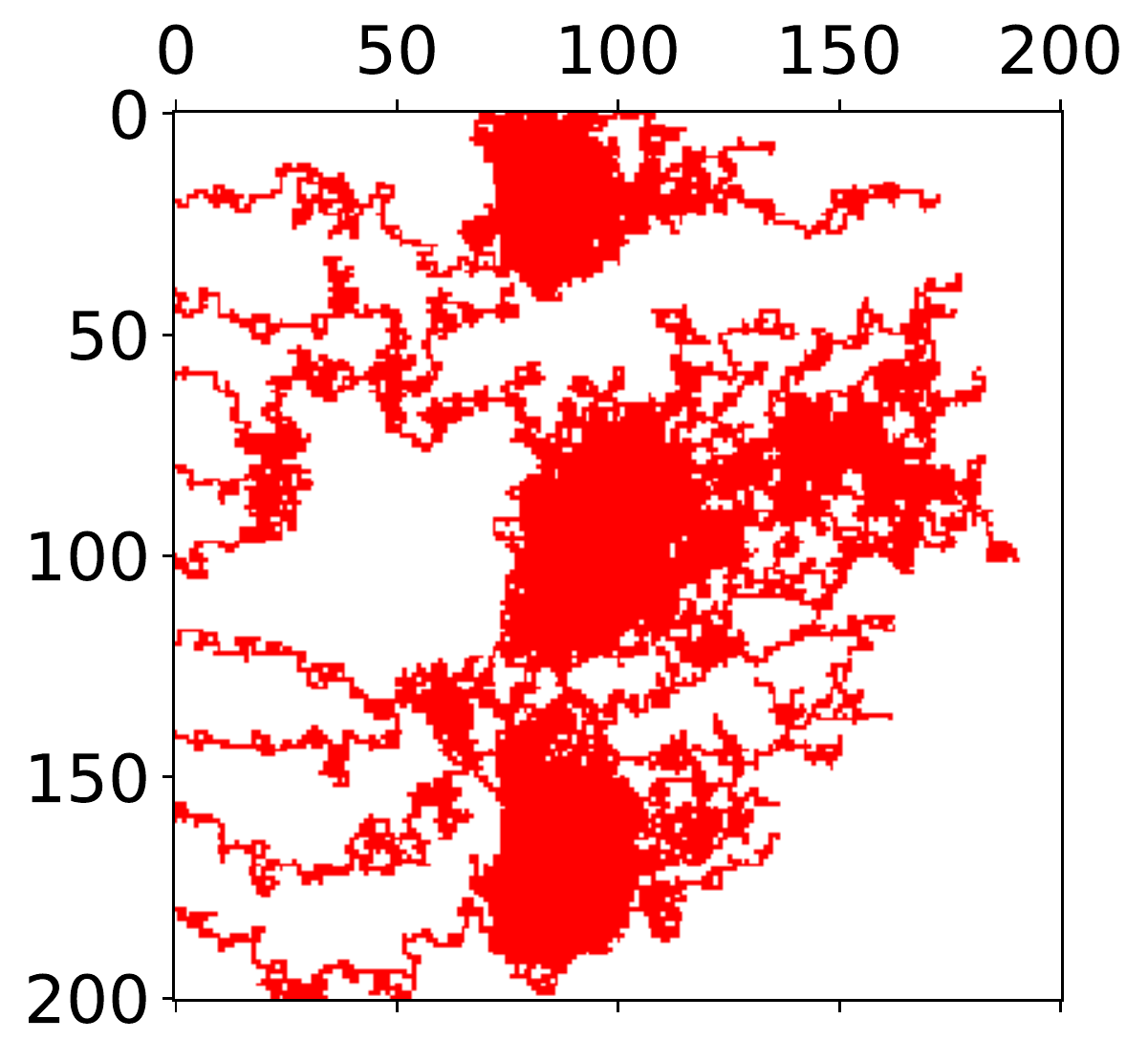}&
\includegraphics[width=0.11\textwidth]{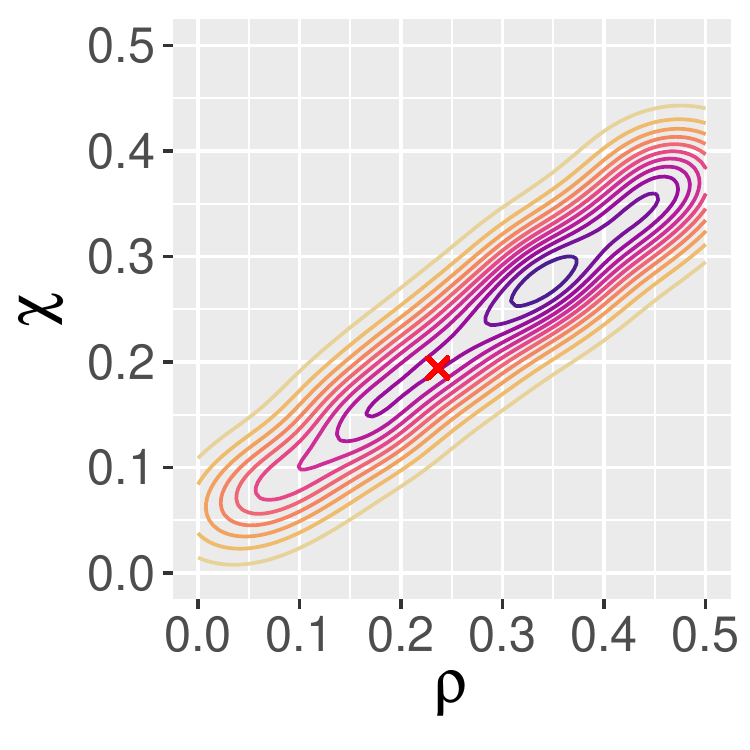}&
\includegraphics[width=0.12\textwidth]{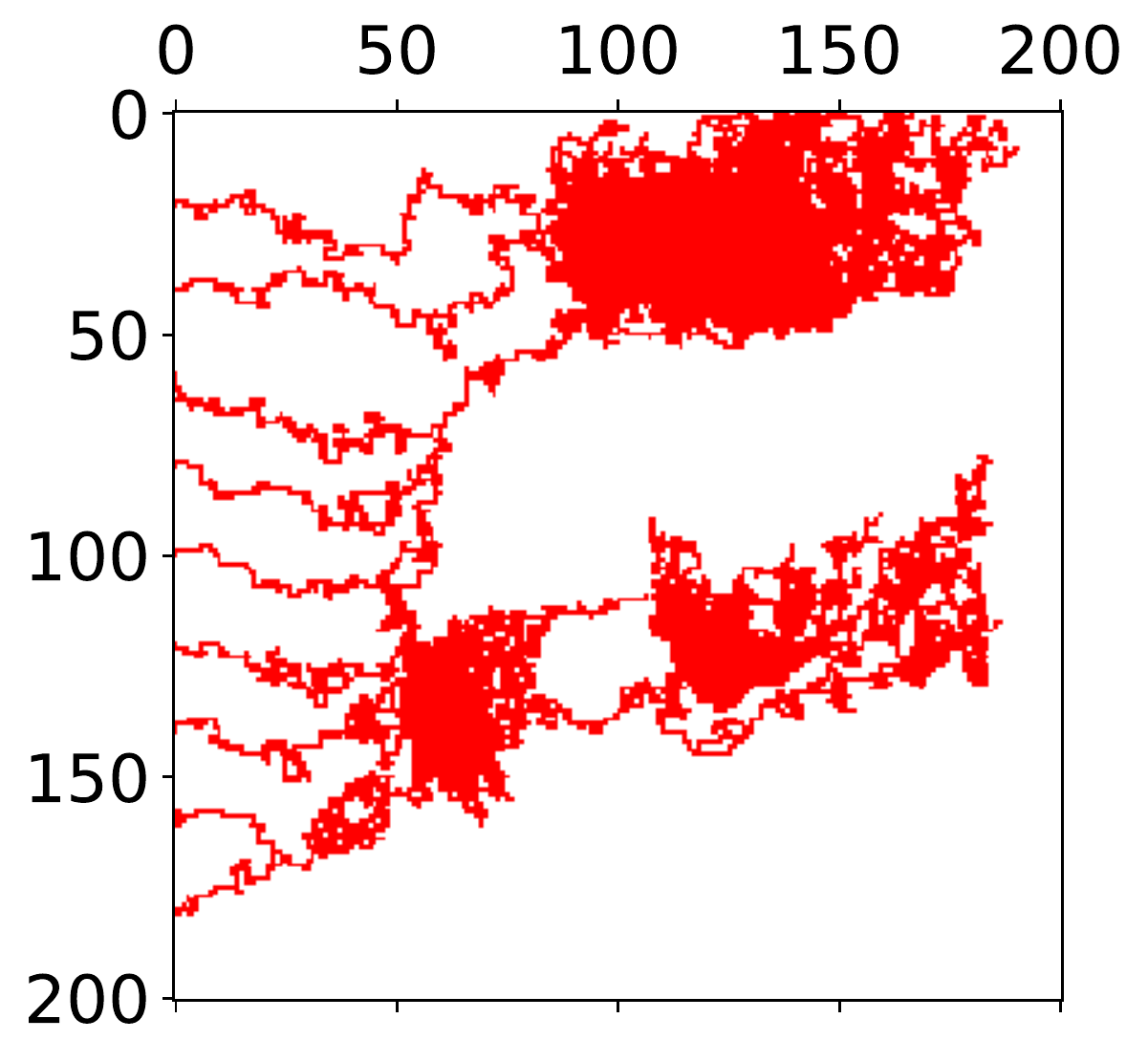}&
\includegraphics[width=0.12\textwidth]{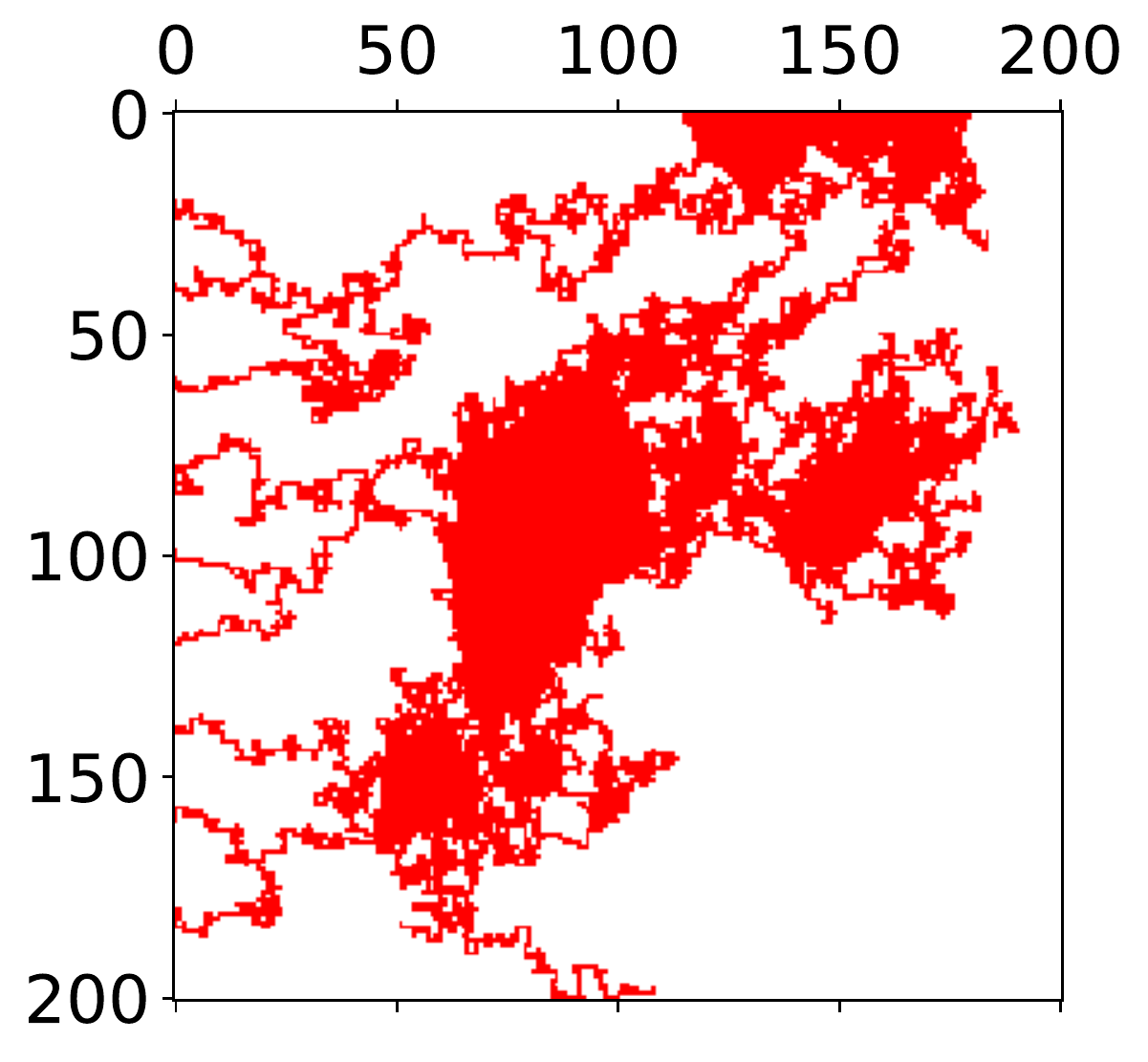}&
\includegraphics[width=0.12\textwidth]{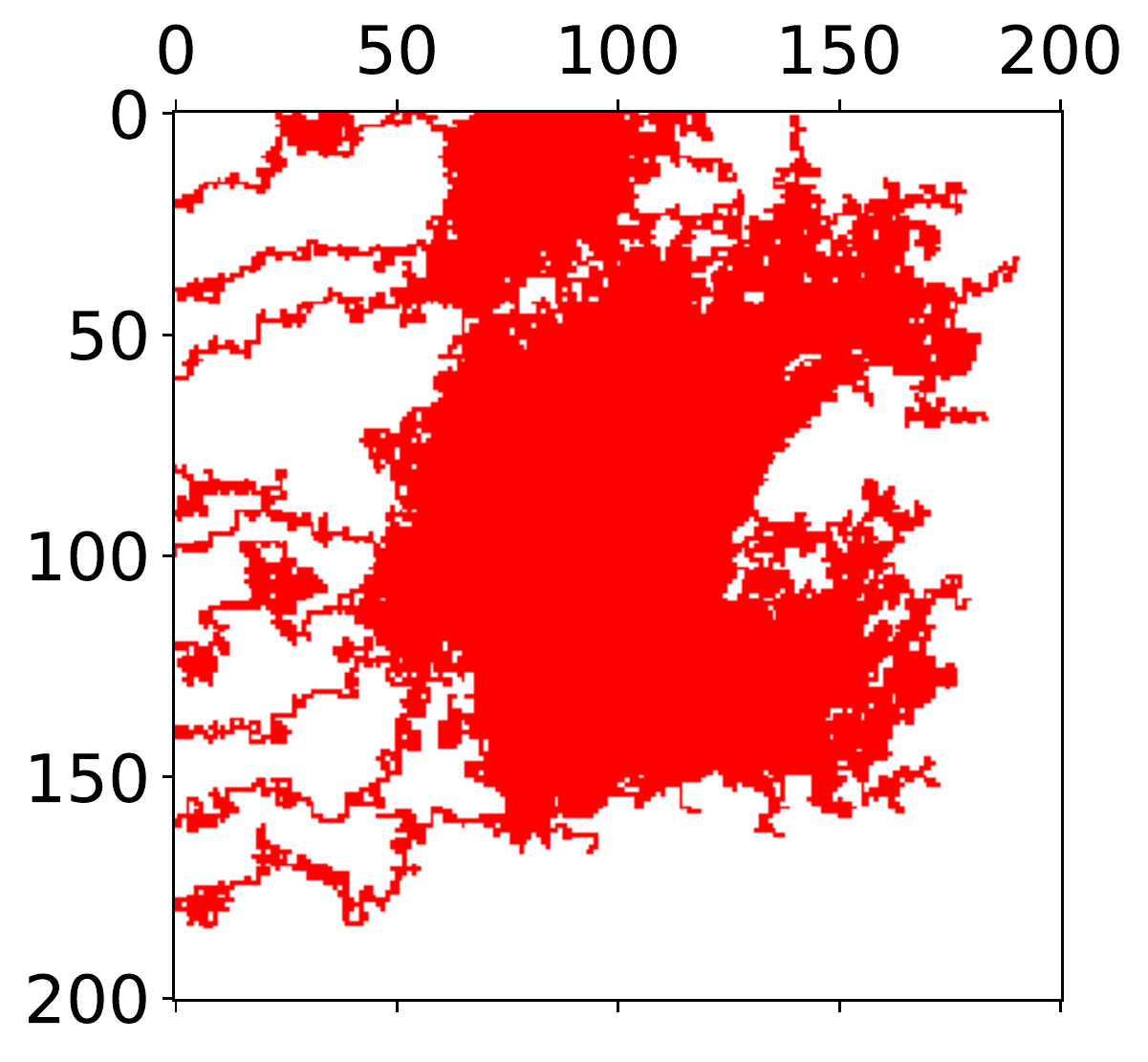}\\

\includegraphics[width=0.12\textwidth]{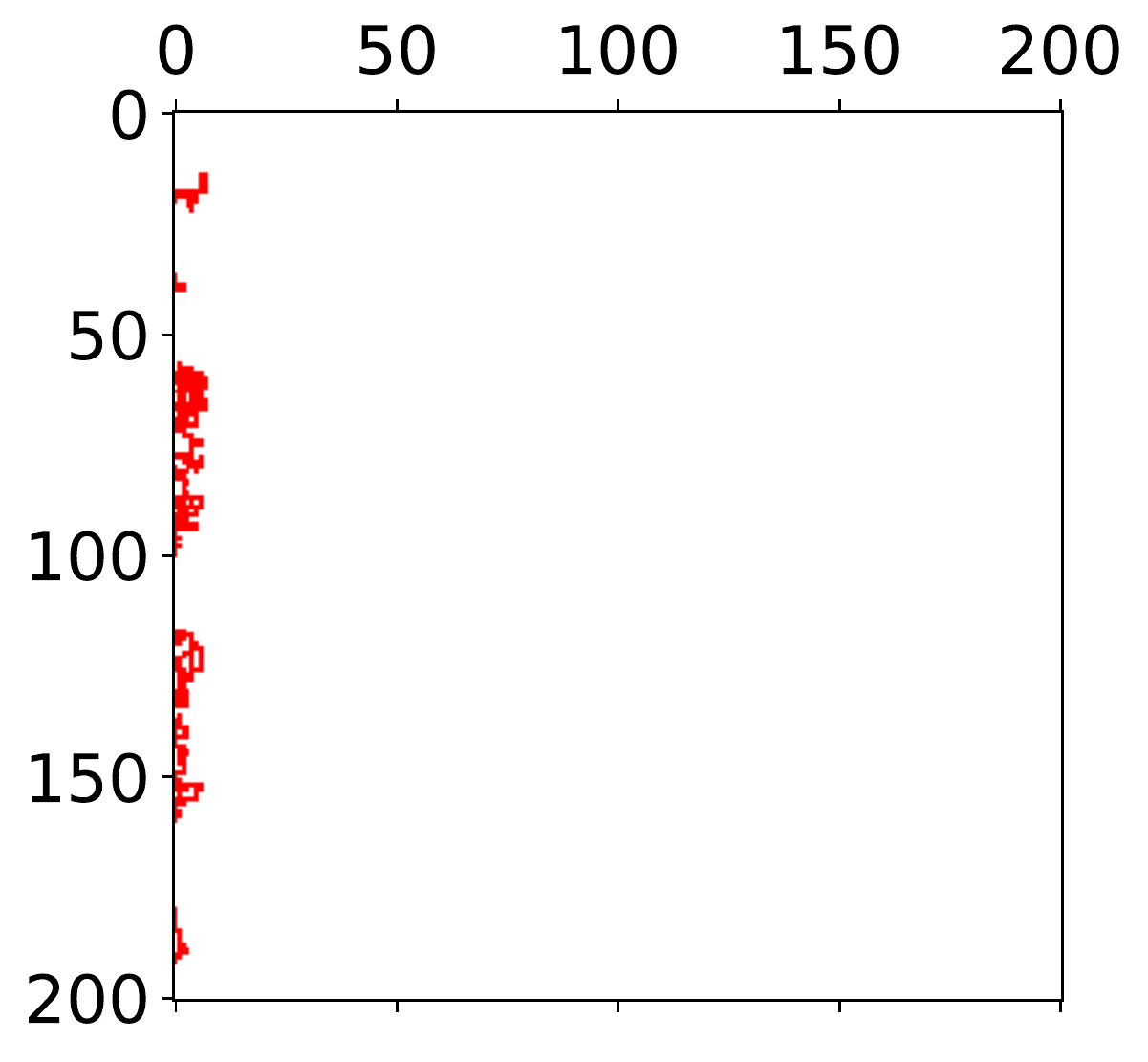}&
\includegraphics[width=0.11\textwidth]{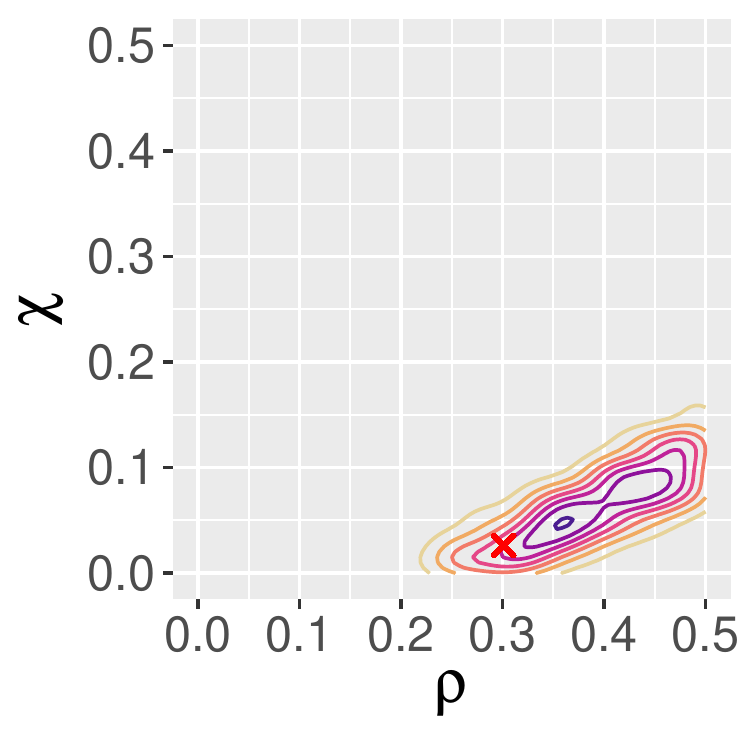}&
\includegraphics[width=0.12\textwidth]{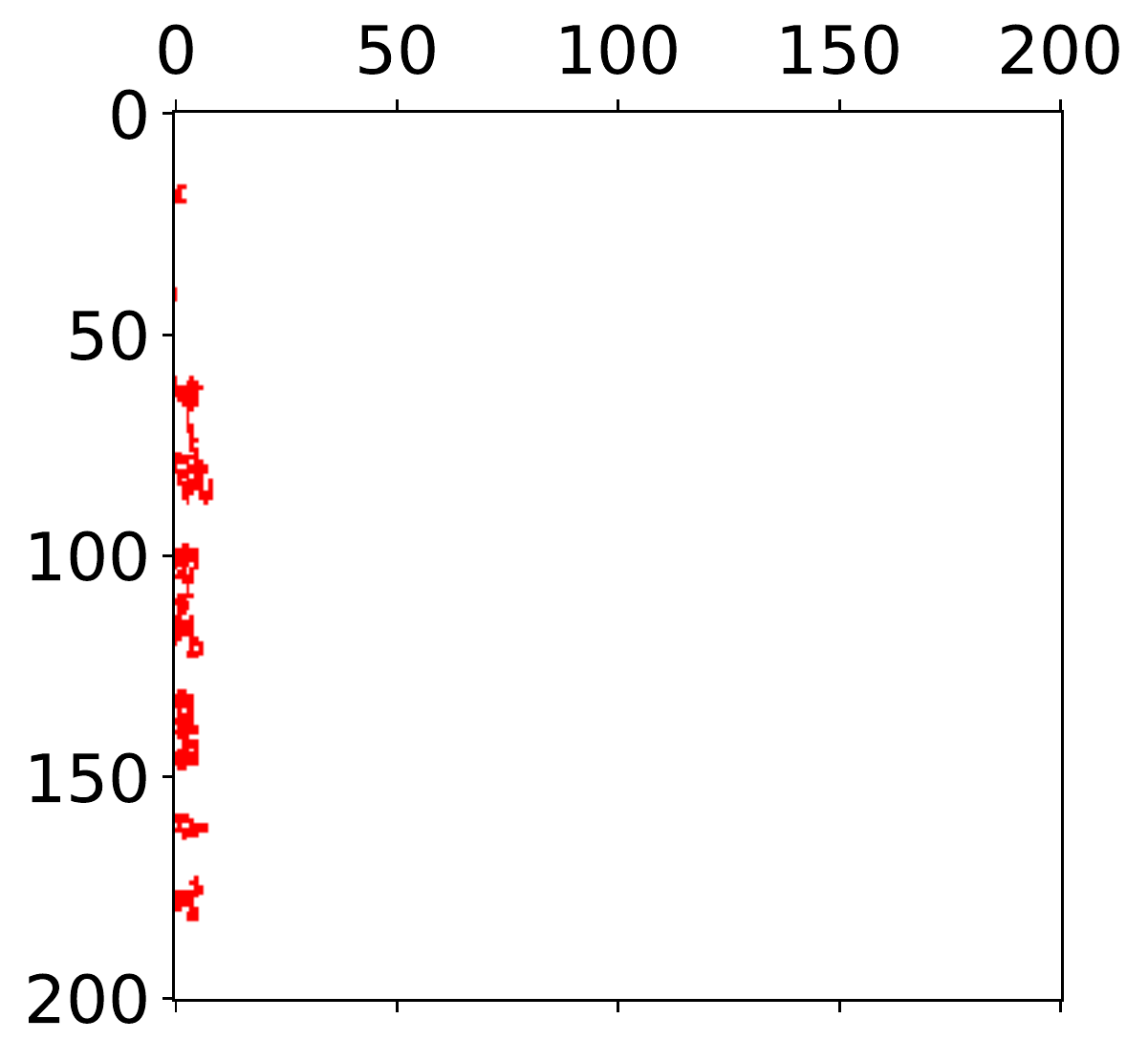}&
\includegraphics[width=0.12\textwidth]{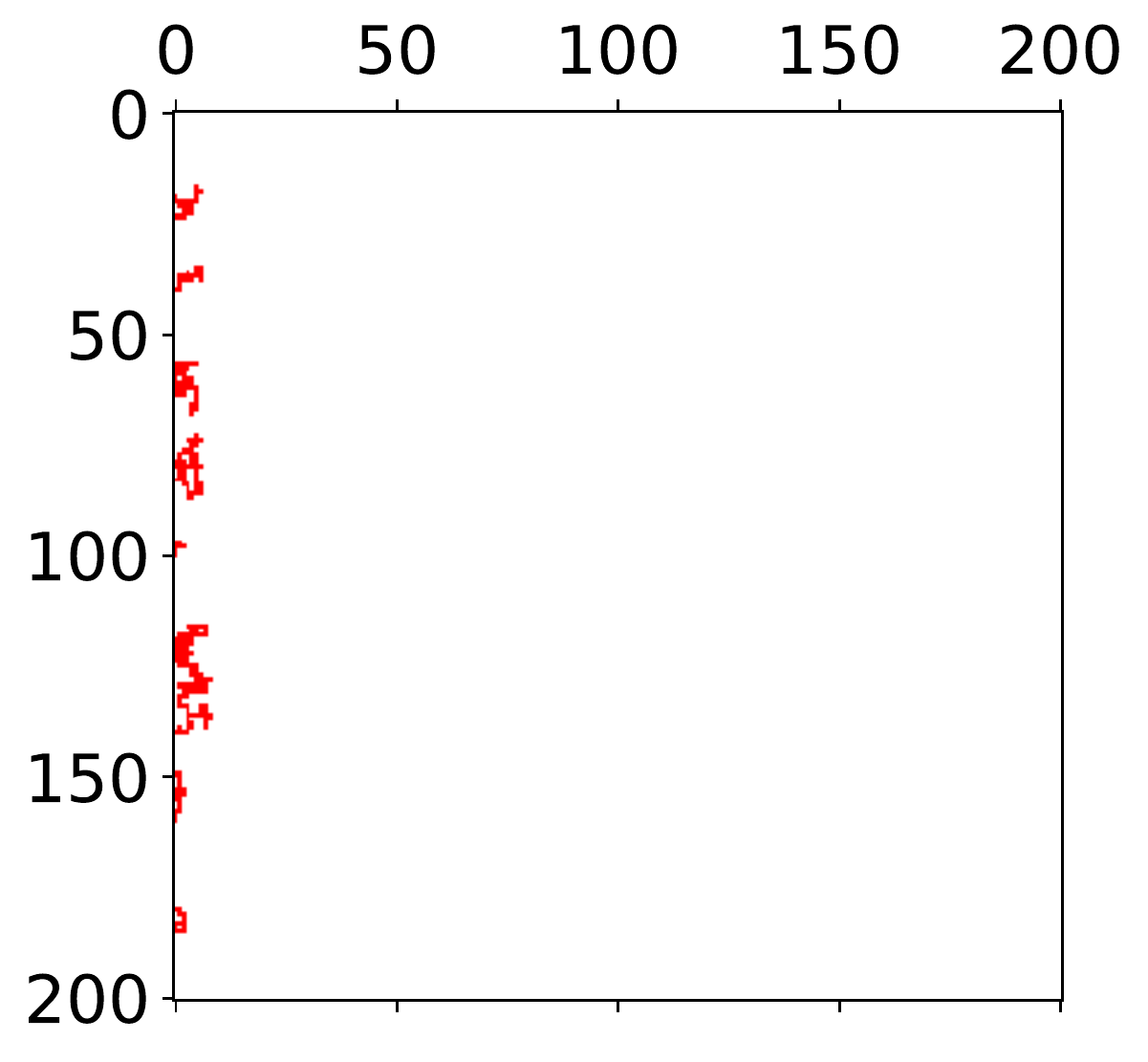}&
\includegraphics[width=0.12\textwidth]{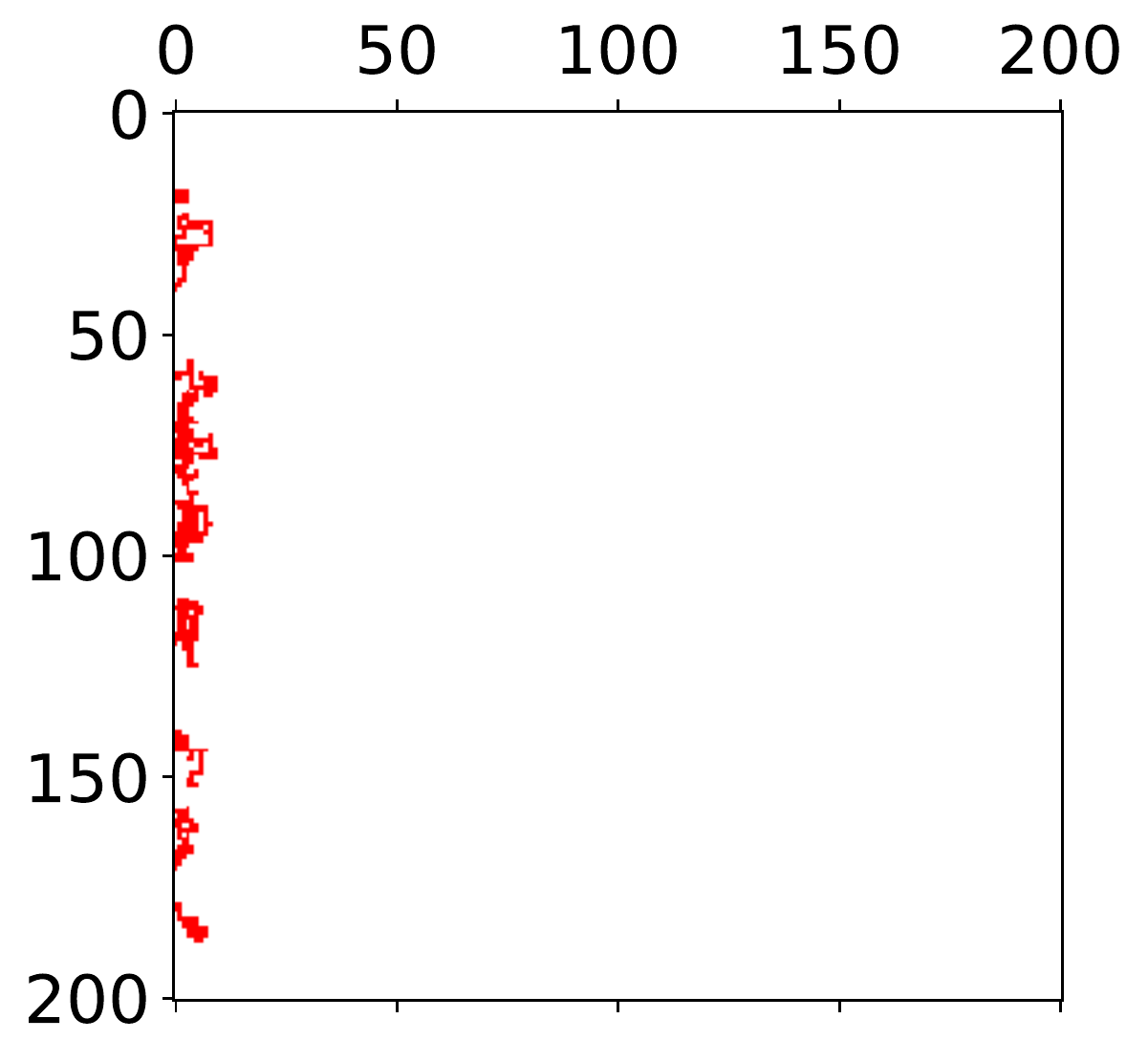}\\

\includegraphics[width=0.12\textwidth]{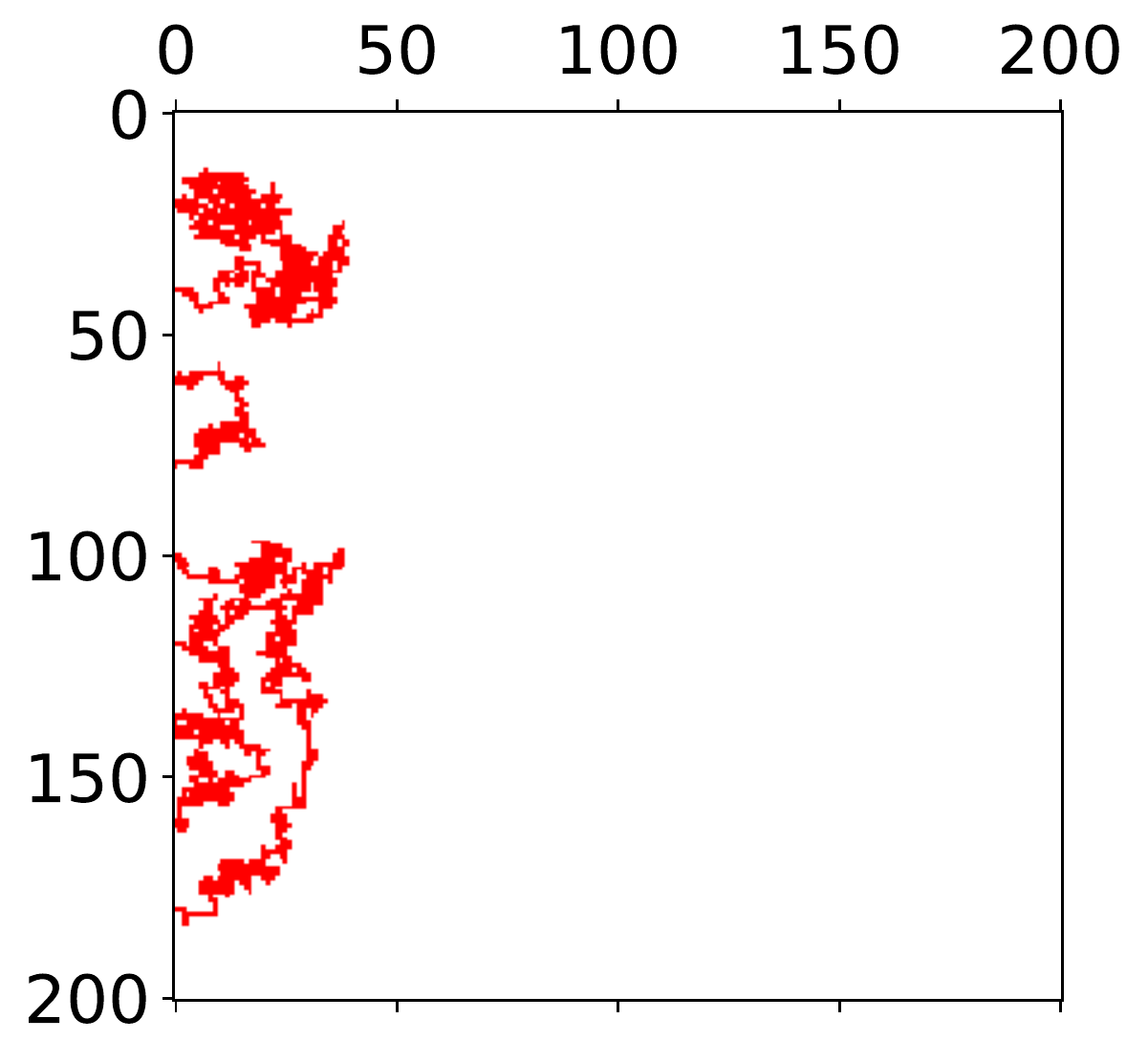}&
\includegraphics[width=0.11\textwidth]{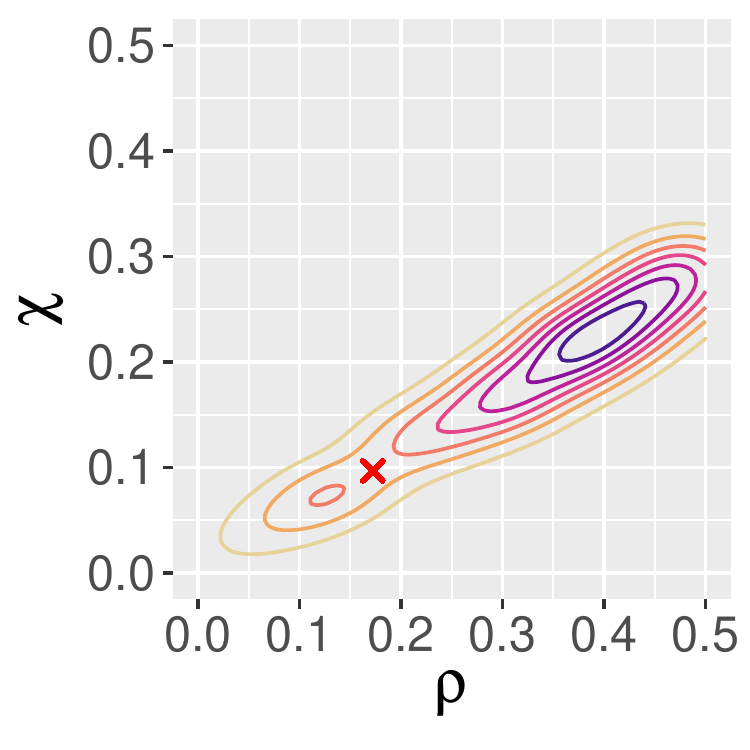}&
\includegraphics[width=0.12\textwidth]{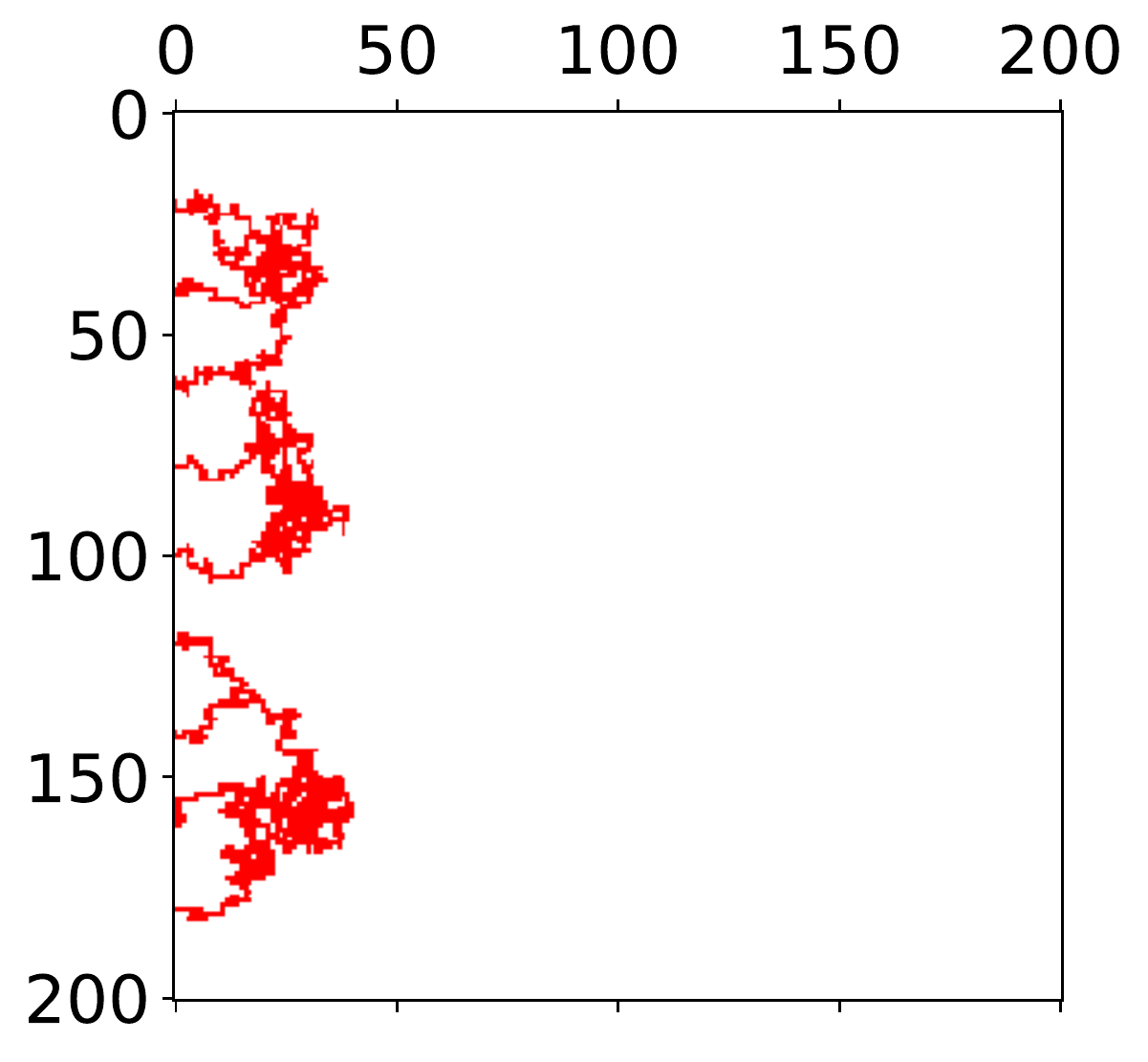}&
\includegraphics[width=0.12\textwidth]{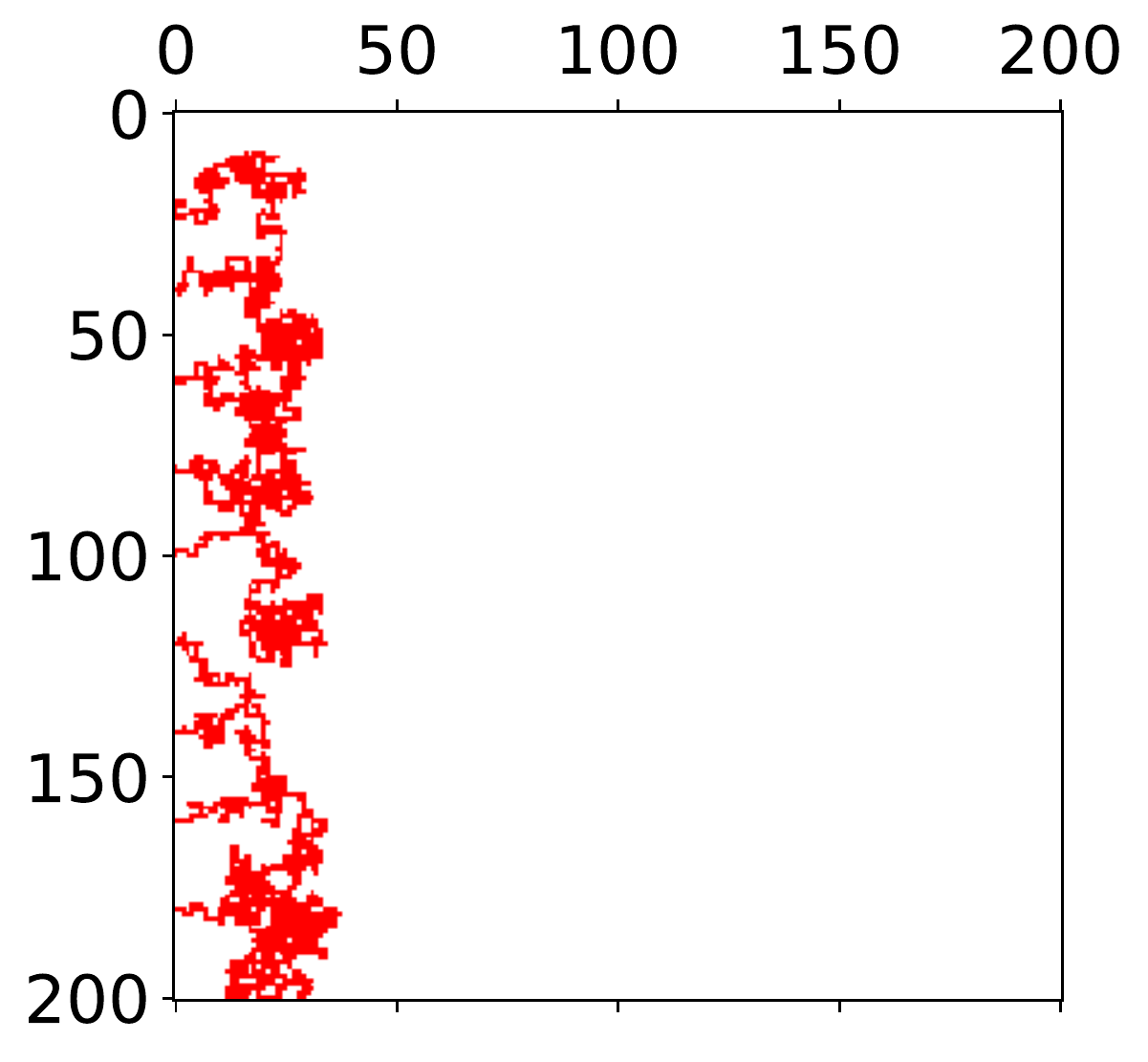}&
\includegraphics[width=0.12\textwidth]{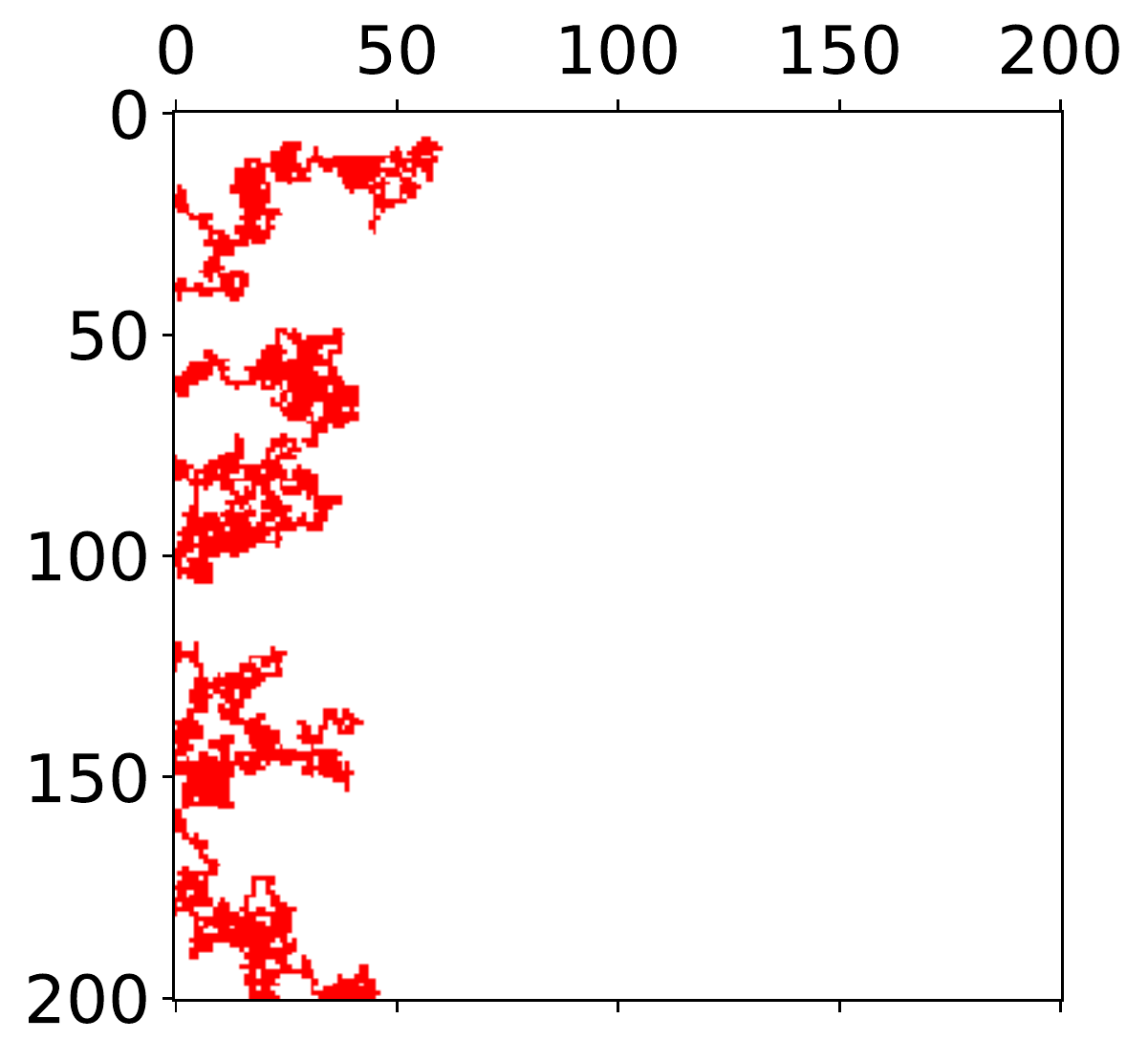}\\
\end{tabular}
\caption{Visualisations of simulation output from the Anderson-Chaplain model for five parameter sets sampled from a uniform prior on the model parameters. The first column shows the observed data, while the second shows a contour plot of the posterior density inferred by applying the TABC methodology, with the red cross indicating the known parameter values used to generate the observed data. The remaining three columns show simulations of parameter values drawn from the ABC posterior predictive distribution.\label{fig:posterior}}
\end{figure*}


We apply the TABC approach described above to parameter inference in the Anderson-Chaplain model. Taking $10000$ samples from the prior on the two model parameters we simulated the Anderson-Chaplain model of angiogenesis for each sampled parameter pair. 


To validate our approach we drew a further $100$ parameter sets from the model prior and simulated data from each to take on the role of the observed data. A representative subset of these simulated data sets can be seen in figure \ref{fig:posterior}, and cover a range of different behaviours.

Given these data, we applied the TABC approach described above to derive samples of $500$ parameter values from the ABC posterior. To investigate the ability of our topological approach to accurately capture the relevant behaviour of the model, we generated ABC posterior predictive samples by simulating the model using parameter values drawn at random from the ABC posterior. These are shown in figure \ref{fig:posterior}, and demonstrate that TABC enables the effective recovery of parameters that replicate the qualitative behaviour of the observed data.

It can be seen that the ABC posterior distributions for the two parameters demonstrate a degree of unidentifiability, in that in most cases the posterior follows a ridge shape with a strong correlation between the two parameters. This aligns with the results found in \citet{nardiniTopologicalDataAnalysis2021}, where it was discovered that there were distinct classes of behaviour that occupied diagonal sections of the parameter space, as do our posterior distributions. Being able to identify such uncertainty in our parameter estimates is one of the key benefits of a Bayesian analysis, and it also provides insights into the behaviour of the model. For example we can see that ABC posterior predictive samples in figure \ref{fig:posterior} are representative of a given class of model behaviour, and that draws from across the potentially wide distribution of parameters indicated by the posterior will follow this behaviour.

The known parameter values used to generate the data on which the posterior distributions are based are marked in figure \ref{fig:posterior}, and can be seen to be within the bulk of the ABC posterior mass.

\begin{table}[h]
\centering
\begin{tabular}{|c|c|c|c|c|}
\hline
Statistics & Mean RSSE & $2\sigma_{\overline{x}}$ RSSE & Mean Entropy & $2\sigma_{\overline{x}}$ Entropy\\
\hline
Image & 4.30 & 0.25 & -2.86 & 0.12\\
Topological & 3.61 & 0.27 & -3.31 & 0.12\\
\hline
\end{tabular}
\vspace*{5pt}
\caption{Mean of the root sum of squared errors and entropy of the posterior distribution inferred from simulated data for $100$ parameter sets drawn from a uniform prior. Values for both the TABC based posterior and ABC on the image-based statistics are shown.}
\label{t:rsse} 
\end{table}

To further quantify the efficacy of our approach, we compared statistics of the posterior distributions obtained from TABC with those generated by an ABC approach using only the image-based statistics described in section \ref{s:im}. We quantified the accuracy of the inferred parameters by taking the mean root sum of squared errors (RSSE) between the posterior samples and the ``true" parameters used to generate the data, as shown in table \ref{t:rsse}. Here the mean RSSE achieved by the topological posterior over the $100$ simulated data sets is below that of the posterior generated using image-based statistics.  We also calculated the mean entropy of the posterior distributions produced for each observed data point using both TABC, and ABC with image-based statistics. As can be seen in table \ref{t:rsse}, the entropy for the posterior derived from the topological features is lower than that derived from the image-based statistics. Taken together, the RSSE and entropy results suggest that the topological statistics used in TABC retain more of the information in the original data set, and hence that TABC is able to more accurately infer the parameters used to generate the data, than ABC using image-based statistics alone.

\section{Conclusions}


We have developed an approach for performing ABC in a topological context that is able to derive posterior distributions over model parameters that can accurately reproduce multiple different classes of behaviour and structure observed within the data. 
We applied extended persistence, which strictly quantifies more topological features than ordinary persistence. Other topological shape statistics have focussed on sweeping across data in multiple different directions  \citep{turner2014persistent,curry2018many,crawford2020predicting}.  Their utility for parameter inference and model selection  will be explored in future studies.

Evaluating the ABC posterior distributions we obtain, we find that by considering topological features in the data through the TABC approach we are able to reduce the posterior uncertainty in the parameter values, and to infer posterior distributions that are more closely focused around the parameters used to generate the data.

While we use persistence images here, there are other potential approaches to summarising topological data analysis for use in parameter inference. For example it is possible to directly derive distances between persistence diagrams in a number of ways \citep{atienzaStabilityPersistentEntropy2020,bubenikStatisticalTopologicalData2015,carriereSlicedWassersteinKernel2017,carriereStableTopologicalSignatures2015,chazalStochasticConvergencePersistence2014,difabioComparingPersistenceDiagrams2015,kerberGeometryHelpsCompare2017,lacombeLargeScaleComputation2018,royerATOLMeasureVectorization2021}, and these could be substituted for the Euclidean distance between the vectors of persistence images that we apply. In future work we will investigate the possibility of applying a distance function on persistence diagrams in the ABC likelihood and how this influences the efficiency of the algorithm.

For simplicity we have also only considered the simplest form of the ABC algorithm -- many other increasingly sophisticated approaches exist, including Markov Chain Monte Carlo algorithms, Sequential Monte Carlo methods \citep{sissonSequentialMonteCarlo2007a} and rare event schemes \citep{prangleRareEventApproach2018}. It would be expected that for models with larger numbers of parameters, significant improvements in efficiency could be obtained by applying one of these approaches rather than a rejection sampler based ABC approach. Doing so would not require any changes to the topological aspects of TABC, only the encompassing sampling mechanism.

As with some other applications of ABC \citep[e.g.][]{russell2019bayesian}, a potential strength of our approach is that it enables a form of {\em qualitative} inference to be performed; in our case by allowing combinations of parameters that result in model behaviour that is topologically similar to the observed data to be identified.  Although we consider a specific application, to parameter inference in the Anderson-Chaplain model of angiogenesis, the TABC approach may be adapted to be widely applicable to parametric models having topological features in the data that are informative about model parameters, including in situations where a mixture of topological statistics and other complementary statistics could be used.  

\section*{Acknowledgements}
HAH thanks PG Kevrekidis and N Whitaker for first presenting the challenge of inferring parameters in models of angiogenesis. HAH also thanks members of the Centre for TDA, specifically A Barbensi, H Byrne, L Marsh, and U Tillmann for many stimulating discussions and helpful comments.  PDWK is grateful to L Reali for useful conversations.

\section*{Funding}
PDWK acknowledges the Medical Research Council (MC\_UU\_00002/13), and support from the National Institute for Health Research (Cambridge Biomedical Research Centre at the Cambridge University Hospitals NHS Foundation Trust). The views expressed are those of the authors and not necessarily those of the NHS, the NIHR, or the Department of Health and Social Care. HAH gratefully acknowledges funding from EPSRC EP/R018472/1, EP/R005125/1 and EP/T001968/1, the Royal Society RGF$\backslash$EA$\backslash$201074 and UF150238, and Emerson Collective.  Partly funded by the RESCUER project. RESCUER has received funding from the European Union's Horizon 2020 research and innovation programme under grant agreement No. 847912.

\section*{Data Availability Statement}

All code used to produce our results is available as a Snakemake \citep{kosterSnakemakeScalableBioinformatics2012} workflow from \url{github.com/tt104/tabc_angio}. It is also stored as an archive on Zenodo with DOI 10.5281/zenodo.5562670.

\bibliographystyle{natbib}
\bibliography{abctda}

\begin{thebibliography}{}

\bibitem[Adams {\em et~al.}(2017)Adams, Emerson, Kirby, Neville, Peterson,
  Shipman, Chepushtanova, Hanson, Motta, and Ziegelmeier]{JMLR:v18:16-337}
Adams, H., Emerson, T., Kirby, M., Neville, R., Peterson, C., Shipman, P.,
  Chepushtanova, S., Hanson, E., Motta, F., and Ziegelmeier, L. (2017).
\newblock Persistence images: {{A}} stable vector representation of persistent
  homology.
\newblock {\em Journal of Machine Learning Research\/}, {\bf 18}(8), 1--35.

\bibitem[Agarwal {\em et~al.}(2006)Agarwal, Edelsbrunner, Harer, and
  Wang]{agarwal2006extreme}
Agarwal, P.~K., Edelsbrunner, H., Harer, J., and Wang, Y. (2006).
\newblock Extreme elevation on a 2-manifold.
\newblock {\em Discrete \& Computational Geometry\/}, {\bf 36}(4), 553--572.

\bibitem[Anderson and Chaplain(1998)Anderson and
  Chaplain]{andersonContinuousDiscreteMathematical1998}
Anderson, A.~R. and Chaplain, M.~A. (1998).
\newblock Continuous and discrete mathematical models of tumor-induced
  angiogenesis.
\newblock {\em Bulletin of Mathematical Biology\/}, {\bf 60}(5), 857--899.

\bibitem[Atienza {\em et~al.}(2020)Atienza, {Gonzalez-D{\'i}az}, and
  {Soriano-Trigueros}]{atienzaStabilityPersistentEntropy2020}
Atienza, N., {Gonzalez-D{\'i}az}, R., and {Soriano-Trigueros}, M. (2020).
\newblock On the stability of persistent entropy and new summary functions for
  topological data analysis.
\newblock {\em Pattern Recognition\/}, {\bf 107}, 107509.

\bibitem[Beaumont {\em et~al.}(2002)Beaumont, Zhang, and
  Balding]{beaumontApproximateBayesianComputation2002}
Beaumont, M.~A., Zhang, W., and Balding, D.~J. (2002).
\newblock Approximate {{Bayesian Computation}} in {{Population Genetics}}.
\newblock {\em Genetics\/}, {\bf 162}(4), 2025--2035.

\bibitem[Beaumont {\em et~al.}(2009)Beaumont, Cornuet, Marin, and
  Robert]{beaumontAdaptiveApproximateBayesian2009}
Beaumont, M.~A., Cornuet, J.-M., Marin, J.-M., and Robert, C.~P. (2009).
\newblock Adaptive approximate {{Bayesian}} computation.
\newblock {\em Biometrika\/}, {\bf 96}(4), 983--990.

\bibitem[Bendich {\em et~al.}(2016)Bendich, Marron, Miller, Pieloch, and
  Skwerer]{bendich2016persistent}
Bendich, P., Marron, J.~S., Miller, E., Pieloch, A., and Skwerer, S. (2016).
\newblock Persistent homology analysis of brain artery trees.
\newblock {\em The annals of applied statistics\/}, {\bf 10}(1), 198.

\bibitem[Bubenik(2015)Bubenik]{bubenikStatisticalTopologicalData2015}
Bubenik, P. (2015).
\newblock Statistical {{Topological Data Analysis}} using {{Persistence
  Landscapes}}.
\newblock {\em Journal of Machine Learning Research\/}, {\bf 16}(3), 77--102.

\bibitem[Carlsson(2009)Carlsson]{carlssonTopologyData2009a}
Carlsson, G. (2009).
\newblock Topology and data.
\newblock {\em Bulletin of the American Mathematical Society\/}, {\bf 46}(2),
  255--308.

\bibitem[Carri{\`e}re {\em et~al.}(2015)Carri{\`e}re, Oudot, and
  Ovsjanikov]{carriereStableTopologicalSignatures2015}
Carri{\`e}re, M., Oudot, S.~Y., and Ovsjanikov, M. (2015).
\newblock Stable {{Topological Signatures}} for {{Points}} on {{3D Shapes}}.

\bibitem[Carri{\`e}re {\em et~al.}(2017)Carri{\`e}re, Cuturi, and
  Oudot]{carriereSlicedWassersteinKernel2017}
Carri{\`e}re, M., Cuturi, M., and Oudot, S. (2017).
\newblock Sliced {{Wasserstein Kernel}} for {{Persistence Diagrams}}.
\newblock In {\em International {{Conference}} on {{Machine Learning}}\/},
  pages 664--673. {PMLR}.

\bibitem[Chazal {\em et~al.}(2014)Chazal, Fasy, Lecci, Rinaldo, and
  Wasserman]{chazalStochasticConvergencePersistence2014}
Chazal, F., Fasy, B.~T., Lecci, F., Rinaldo, A., and Wasserman, L. (2014).
\newblock Stochastic {{Convergence}} of {{Persistence Landscapes}} and
  {{Silhouettes}}.
\newblock In {\em Proceedings of the Thirtieth Annual Symposium on
  {{Computational}} Geometry\/}, {{SOCG}}'14, pages 474--483, {New York, NY,
  USA}. {Association for Computing Machinery}.

\bibitem[{Cohen-Steiner} {\em et~al.}(2009){Cohen-Steiner}, Edelsbrunner, and
  Harer]{cohen-steinerExtendingPersistenceUsing2009}
{Cohen-Steiner}, D., Edelsbrunner, H., and Harer, J. (2009).
\newblock Extending {{Persistence Using Poincar\'e}} and {{Lefschetz Duality}}.
\newblock {\em Foundations of Computational Mathematics\/}, {\bf 9}(1),
  79--103.

\bibitem[Crawford {\em et~al.}(2020)Crawford, Monod, Chen, Mukherjee, and
  Rabad{\'a}n]{crawford2020predicting}
Crawford, L., Monod, A., Chen, A.~X., Mukherjee, S., and Rabad{\'a}n, R.
  (2020).
\newblock Predicting clinical outcomes in glioblastoma: an application of
  topological and functional data analysis.
\newblock {\em Journal of the American Statistical Association\/}, {\bf
  115}(531), 1139--1150.

\bibitem[Curry {\em et~al.}(2018)Curry, Mukherjee, and Turner]{curry2018many}
Curry, J., Mukherjee, S., and Turner, K. (2018).
\newblock How many directions determine a shape and other sufficiency results
  for two topological transforms.
\newblock {\em arXiv preprint arXiv:1805.09782\/}.

\bibitem[Del~Moral {\em et~al.}(2012)Del~Moral, Doucet, and
  Jasra]{delmoralAdaptiveSequentialMonte2012}
Del~Moral, P., Doucet, A., and Jasra, A. (2012).
\newblock An adaptive sequential {{Monte Carlo}} method for approximate
  {{Bayesian}} computation.
\newblock {\em Statistics and Computing\/}, {\bf 22}(5), 1009--1020.

\bibitem[Di~Fabio and Ferri(2015)Di~Fabio and
  Ferri]{difabioComparingPersistenceDiagrams2015}
Di~Fabio, B. and Ferri, M. (2015).
\newblock Comparing {{Persistence Diagrams Through Complex Vectors}}.
\newblock In V.~Murino and E.~Puppo, editors, {\em Image {{Analysis}} and
  {{Processing}} \textemdash{} {{ICIAP}} 2015\/}, Lecture {{Notes}} in
  {{Computer Science}}, pages 294--305, {Cham}. {Springer International
  Publishing}.

\bibitem[Edelsbrunner and Harer(2010)Edelsbrunner and
  Harer]{edelsbrunnerComputationalTopologyIntroduction2010}
Edelsbrunner, H. and Harer, J. (2010).
\newblock {\em Computational {{Topology}}: {{An Introduction}}\/}.
\newblock {American Mathematical Soc.}

\bibitem[Fu and Li(1997)Fu and Li]{fuEstimatingAgeCommon1997}
Fu, Y.~X. and Li, W.~H. (1997).
\newblock Estimating the age of the common ancestor of a sample of {{DNA}}
  sequences.
\newblock {\em Molecular Biology and Evolution\/}, {\bf 14}(2), 195--199.

\bibitem[Ghrist(2018)Ghrist]{ghristHomologicalAlgebraData2018a}
Ghrist, R. (2018).
\newblock Homological algebra and data.
\newblock In {\em The {{Mathematics}} of {{Data}}\/}, volume~25 of {\em
  {{IAS}}/{{Park City Mathematics Series}}\/}. {American Mathematical Society}.

\bibitem[Kerber {\em et~al.}(2017)Kerber, Morozov, and
  Nigmetov]{kerberGeometryHelpsCompare2017}
Kerber, M., Morozov, D., and Nigmetov, A. (2017).
\newblock Geometry {{Helps}} to {{Compare Persistence Diagrams}}.
\newblock {\em ACM Journal of Experimental Algorithmics\/}, {\bf 22},
  1.4:1--1.4:20.

\bibitem[Kirk {\em et~al.}(2015)Kirk, Babtie, and Stumpf]{kirk2015systems}
Kirk, P., Babtie, A.~C., and Stumpf, M.~P. (2015).
\newblock Systems biology (un)certainties.
\newblock {\em Science\/}, {\bf 350}(6259), 386--388.

\bibitem[K{\"o}ster and Rahmann(2012)K{\"o}ster and
  Rahmann]{kosterSnakemakeScalableBioinformatics2012}
K{\"o}ster, J. and Rahmann, S. (2012).
\newblock Snakemake\textemdash a scalable bioinformatics workflow engine.
\newblock {\em Bioinformatics\/}, {\bf 28}(19), 2520--2522.

\bibitem[Lacombe {\em et~al.}(2018)Lacombe, Cuturi, and
  OUDOT]{lacombeLargeScaleComputation2018}
Lacombe, T., Cuturi, M., and OUDOT, S. (2018).
\newblock Large {{Scale}} computation of {{Means}} and {{Clusters}} for
  {{Persistence Diagrams}} using {{Optimal Transport}}.
\newblock In {\em Advances in {{Neural Information Processing Systems}}\/},
  volume~31. {Curran Associates, Inc.}

\bibitem[Liepe {\em et~al.}(2014)Liepe, Kirk, Filippi, Toni, Barnes, and
  Stumpf]{liepe2014framework}
Liepe, J., Kirk, P., Filippi, S., Toni, T., Barnes, C.~P., and Stumpf, M.~P.
  (2014).
\newblock A framework for parameter estimation and model selection from
  experimental data in systems biology using approximate bayesian computation.
\newblock {\em Nature protocols\/}, {\bf 9}(2), 439--456.

\bibitem[Marjoram {\em et~al.}(2003)Marjoram, Molitor, Plagnol, and
  Tavar{\'e}]{marjoramMarkovChainMonte2003a}
Marjoram, P., Molitor, J., Plagnol, V., and Tavar{\'e}, S. (2003).
\newblock Markov chain {{Monte Carlo}} without likelihoods.
\newblock {\em Proceedings of the National Academy of Sciences\/}, {\bf
  100}(26), 15324--15328.

\bibitem[Maroulas {\em et~al.}(2020)Maroulas, Nasrin, and
  Oballe]{maroulasBayesianFrameworkPersistent2020}
Maroulas, V., Nasrin, F., and Oballe, C. (2020).
\newblock A {{Bayesian Framework}} for {{Persistent Homology}}.
\newblock {\em SIAM Journal on Mathematics of Data Science\/}, {\bf 2}(1),
  48--74.

\bibitem[McGuirl {\em et~al.}(2020)McGuirl, Volkening, and
  Sandstede]{mcguirl2020topological}
McGuirl, M.~R., Volkening, A., and Sandstede, B. (2020).
\newblock Topological data analysis of zebrafish patterns.
\newblock {\em Proceedings of the National Academy of Sciences\/}, {\bf
  117}(10), 5113--5124.

\bibitem[Murray(2003)Murray]{murrayMathematicalBiologyII2003}
Murray, J.~D. (2003).
\newblock {\em Mathematical {{Biology II}}: {{Spatial Models}} and {{Biomedical
  Applications}}\/}.
\newblock Interdisciplinary {{Applied Mathematics}}, {{Mathematical Biology}}.
  {Springer-Verlag}, {New York}, third edition.

\bibitem[Nardini {\em et~al.}(2021)Nardini, Stolz, Flores, Harrington, and
  Byrne]{nardiniTopologicalDataAnalysis2021}
Nardini, J.~T., Stolz, B.~J., Flores, K.~B., Harrington, H.~A., and Byrne,
  H.~M. (2021).
\newblock Topological data analysis distinguishes parameter regimes in the
  {{Anderson}}-{{Chaplain}} model of angiogenesis.
\newblock {\em arXiv:2101.00523 [q-bio]\/}.

\bibitem[Otter {\em et~al.}(2017)Otter, Porter, Tillmann, Grindrod, and
  Harrington]{Otter2017}
Otter, N., Porter, M.~A., Tillmann, U., Grindrod, P., and Harrington, H.~A.
  (2017).
\newblock A roadmap for the computation of persistent homology.
\newblock {\em European Physical Journal -- Data Science\/}, {\bf 6}(17),
  1--38.

\bibitem[Prangle {\em et~al.}(2018)Prangle, Everitt, and
  Kypraios]{prangleRareEventApproach2018}
Prangle, D., Everitt, R.~G., and Kypraios, T. (2018).
\newblock A rare event approach to high-dimensional approximate {{Bayesian}}
  computation.
\newblock {\em Statistics and Computing\/}, {\bf 28}(4), 819--834.

\bibitem[Robins and Turner(2016)Robins and
  Turner]{robinsPrincipalComponentAnalysis2016}
Robins, V. and Turner, K. (2016).
\newblock Principal component analysis of persistent homology rank functions
  with case studies of spatial point patterns, sphere packing and colloids.
\newblock {\em Physica D: Nonlinear Phenomena\/}, {\bf 334}, 99--117.

\bibitem[Royer {\em et~al.}(2021)Royer, Chazal, Levrard, Umeda, and
  Ike]{royerATOLMeasureVectorization2021}
Royer, M., Chazal, F., Levrard, C., Umeda, Y., and Ike, Y. (2021).
\newblock {{ATOL}}: {{Measure Vectorization}} for {{Automatic
  Topologically}}-{{Oriented Learning}}.
\newblock In {\em International {{Conference}} on {{Artificial Intelligence}}
  and {{Statistics}}\/}, pages 1000--1008. {PMLR}.

\bibitem[Russell-Buckland {\em et~al.}(2019)Russell-Buckland, Barnes, and
  Tachtsidis]{russell2019bayesian}
Russell-Buckland, J., Barnes, C.~P., and Tachtsidis, I. (2019).
\newblock A bayesian framework for the analysis of systems biology models of
  the brain.
\newblock {\em PLoS computational biology\/}, {\bf 15}(4), e1006631.

\bibitem[Sgouralis {\em et~al.}(2017)Sgouralis, Nebenf{\"u}hr, and
  Maroulas]{sgouralisBayesianTopologicalFramework2017}
Sgouralis, I., Nebenf{\"u}hr, A., and Maroulas, V. (2017).
\newblock A {{Bayesian Topological Framework}} for the {{Identification}} and
  {{Reconstruction}} of {{Subcellular Motion}}.
\newblock {\em SIAM Journal on Imaging Sciences\/}, {\bf 10}(2), 871--899.

\bibitem[Sisson {\em et~al.}(2007)Sisson, Fan, and
  Tanaka]{sissonSequentialMonteCarlo2007a}
Sisson, S.~A., Fan, Y., and Tanaka, M.~M. (2007).
\newblock Sequential {{Monte Carlo}} without likelihoods.
\newblock {\em Proceedings of the National Academy of Sciences\/}, {\bf
  104}(6), 1760--1765.

\bibitem[Sisson {\em et~al.}(2018)Sisson, Fan, and
  Beaumont]{sissonHandbookApproximateBayesian2018}
Sisson, S.~A., Fan, Y., and Beaumont, M.~A., editors (2018).
\newblock {\em Handbook of {{Approximate Bayesian Computation}}\/}.
\newblock {Chapman and Hall/CRC}, {Boca Raton}.

\bibitem[Stolz {\em et~al.}(2020)Stolz, Kaeppler, Markelc, Mech, Lipsmeier,
  Muschel, Byrne, and Harrington]{stolz2020multiscale}
Stolz, B.~J., Kaeppler, J., Markelc, B., Mech, F., Lipsmeier, F., Muschel,
  R.~J., Byrne, H.~M., and Harrington, H.~A. (2020).
\newblock Multiscale topology characterises dynamic tumour vascular networks.
\newblock {\em arXiv preprint arXiv:2008.08667\/}.

\bibitem[Stolz-Pretzer(2019)Stolz-Pretzer]{stolz-pretzer_global_2019}
Stolz-Pretzer, B. (2019).
\newblock {\em Global and local persistent homology for the shape and
  classification of biological data\/}.
\newblock http://purl.org/dc/dcmitype/{Text}, University of Oxford.

\bibitem[Tavar{\'e} {\em et~al.}(1997)Tavar{\'e}, Balding, Griffiths, and
  Donnelly]{tavareInferringCoalescenceTimes1997}
Tavar{\'e}, S., Balding, D.~J., Griffiths, R.~C., and Donnelly, P. (1997).
\newblock Inferring {{Coalescence Times From DNA Sequence Data}}.
\newblock {\em Genetics\/}, {\bf 145}(2), 505--518.

\bibitem[Thorne and Stumpf(2012)Thorne and
  Stumpf]{thorneGraphSpectralAnalysis2012a}
Thorne, T. and Stumpf, M. P.~H. (2012).
\newblock Graph spectral analysis of protein interaction network evolution.
\newblock {\em Journal of The Royal Society Interface\/}, {\bf 9}(75),
  2653--2666.

\bibitem[Toni {\em et~al.}(2009)Toni, Welch, Strelkowa, Ipsen, and
  Stumpf]{toniApproximateBayesianComputation2009a}
Toni, T., Welch, D., Strelkowa, N., Ipsen, A., and Stumpf, M.~P. (2009).
\newblock Approximate {{Bayesian}} computation scheme for parameter inference
  and model selection in dynamical systems.
\newblock {\em Journal of The Royal Society Interface\/}, {\bf 6}(31),
  187--202.

\bibitem[Turner {\em et~al.}(2014)Turner, Mukherjee, and
  Boyer]{turner2014persistent}
Turner, K., Mukherjee, S., and Boyer, D.~M. (2014).
\newblock Persistent homology transform for modeling shapes and surfaces.
\newblock {\em Information and Inference: A Journal of the IMA\/}, {\bf 3}(4),
  310--344.

\bibitem[Vipond {\em et~al.}(2021)Vipond, Bull, Macklin, Tillmann, Pugh, Byrne,
  and Harrington]{Vipond2021}
Vipond, O., Bull, J.~A., Macklin, P.~S., Tillmann, U., Pugh, C.~W., Byrne,
  H.~M., and Harrington, H.~A. (2021).
\newblock Multiparameter persistent homology landscapes identify spatial
  patterns of immune cells in tumors.
\newblock {\em Proceedings of the National Academy of Sciences (in press)\/}.

\bibitem[Warne {\em et~al.}(2019)Warne, Baker, and Simpson]{warne2019using}
Warne, D.~J., Baker, R.~E., and Simpson, M.~J. (2019).
\newblock Using experimental data and information criteria to guide model
  selection for reaction--diffusion problems in mathematical biology.
\newblock {\em Bulletin of Mathematical Biology\/}, {\bf 81}(6), 1760--1804.

\bibitem[Wasserman(2018)Wasserman]{wassermanTopologicalDataAnalysis2018}
Wasserman, L. (2018).
\newblock Topological {{Data Analysis}}.
\newblock {{SSRN Scholarly Paper}} ID 3156968, {Social Science Research
  Network}, {Rochester, NY}.

\bibitem[Yim and Leygonie(2021)Yim and Leygonie]{yim2021optimization}
Yim, K.~M. and Leygonie, J. (2021).
\newblock Optimization of spectral wavelets for persistence-based graph
  classification.
\newblock {\em Frontiers in Applied Mathematics and Statistics\/}, {\bf 7}, 16.

\end{thebibliography}

\end{document}